%% file: AAAAmain.tex
\documentclass[journal]{IEEEtran}
\usepackage{times}
\usepackage{amssymb}
\usepackage{helvet}
\usepackage{mathrsfs}
\usepackage{url}
\usepackage{courier}
\usepackage{booktabs,caption}
\usepackage{threeparttable}
\usepackage{graphicx}
\usepackage{booktabs}
\usepackage{url}	
\usepackage{overpic}
\usepackage{color}
\usepackage[english]{babel}
\usepackage[T1]{fontenc}
\usepackage[utf8]{inputenc}
\usepackage{multirow}
\usepackage{diagbox}
\usepackage{authblk}
\usepackage[numbers, compress]{natbib}
\usepackage{amsmath}
\usepackage{xspace}
\usepackage{subfig}
\usepackage{amsmath}
\usepackage{algorithm}
\usepackage{algpseudocode}
\usepackage{color}
\usepackage{wrapfig}
\usepackage{multirow}

\makeatletter
\DeclareRobustCommand\onedot{\futurelet\@let@token\@onedot}
\def\@onedot{\ifx\@let@token.\else.\null\fi\xspace}
\def\eg{\emph{e.g}\onedot} 
\def\ie{\emph{i.e}\onedot} 
 
\def\etc{\emph{etc}\onedot} 
 
\def\etal{\emph{et al}\onedot}
\newcommand{\ve}[1]{{\bf #1}}
\makeatother
\newcommand{\para}[1]{\vspace{.05in}\noindent\textbf{#1}}

\begin{document}
\title{CANet: Cross-disease Attention Network  for \\ Joint Diabetic Retinopathy  and  \\Diabetic Macular Edema Grading}

\author{Xiaomeng Li,~\IEEEmembership{Student~Member,~IEEE,}
Xiaowei Hu,
Lequan Yu,~\IEEEmembership{Student~Member,~IEEE,} \\
Lei Zhu,~\IEEEmembership{Member,~IEEE,} 
Chi-Wing Fu,~\IEEEmembership{Member,~IEEE,}
and Pheng-Ann Heng,~\IEEEmembership{Senior~Member,~IEEE}

		
\thanks{X. Li, X. Hu, L. Yu, L. Zhu, C.-W. Fu and P.-A. Heng are with the Department of Computer Science and Engineering, The Chinese University of Hong Kong, Hong Kong (e-mail: {xmli, xwhu, lqyu, lzhu, cwfu, pheng}@cse.cuhk.edu.hk)}
}

\markboth{}
{Shell \MakeLowercase{\textit{et al.}}: Bare Demo of IEEEtran.cls for IEEE Journals}

\newcommand{\revise}[1]{{\color{black}{#1}}}
\newcommand{\revisesecond}[1]{{\color{black}{#1}}}
\newcommand{\xmli}[1]{{\color{blue}{#1}}}
\newcommand{\ylq}[1]{{\color{magenta}{LQ:#1}}}
\newcommand{\xwhu}[1]{{\color{red}{XW:#1}}}
\maketitle

\IEEEpeerreviewmaketitle
\input{abstract}


\input{introduction} 
\input{related}
\input{method}
\input{experiments}

\input{discussion}

\input{conclusion}

\bibliographystyle{IEEEtraN}
\small{\bibliography{refs}}
\ifCLASSOPTIONcaptionsoff
  \newpage
\fi
\end{document}

%% file: abstract.tex
\begin{abstract}
Diabetic retinopathy (DR) and diabetic macular edema (DME) are the leading causes of permanent blindness in the working-age population. 
Automatic grading of DR and DME helps ophthalmologists design tailored treatments to patients, thus is of vital importance in the clinical practice.
However, prior works either grade DR or DME, and ignore the correlation between DR and its complication, \ie, DME. 
Moreover, the location information, \eg, macula and soft hard exhaust annotations, are widely used as a prior for grading. Such annotations are costly to obtain, hence it is desirable to develop automatic grading methods with only image-level supervision.  
In this paper, we present a novel cross-disease attention network (CANet) to jointly grade DR and DME by {\em exploring the internal relationship between the diseases} with only image-level supervision.
Our key contributions include the disease-specific attention module to selectively learn useful features for individual diseases, and the disease-dependent attention module to further capture the internal relationship between the two diseases.
We integrate these two attention modules in a deep network to produce disease-specific and disease-dependent features, and to maximize the overall performance jointly for grading DR and DME. 
We evaluate our network on two public benchmark datasets, \ie, ISBI 2018 IDRiD challenge dataset and Messidor dataset.
Our method achieves the best result on the ISBI 2018 IDRiD challenge dataset and outperforms other methods on the Messidor dataset.
Our code is publicly available at https://github.com/xmengli999/CANet.
\end{abstract}

%% file: introduction.tex
\section{Introduction}
\begin{figure}[t]
	\centering
	\includegraphics[width=0.5\textwidth]{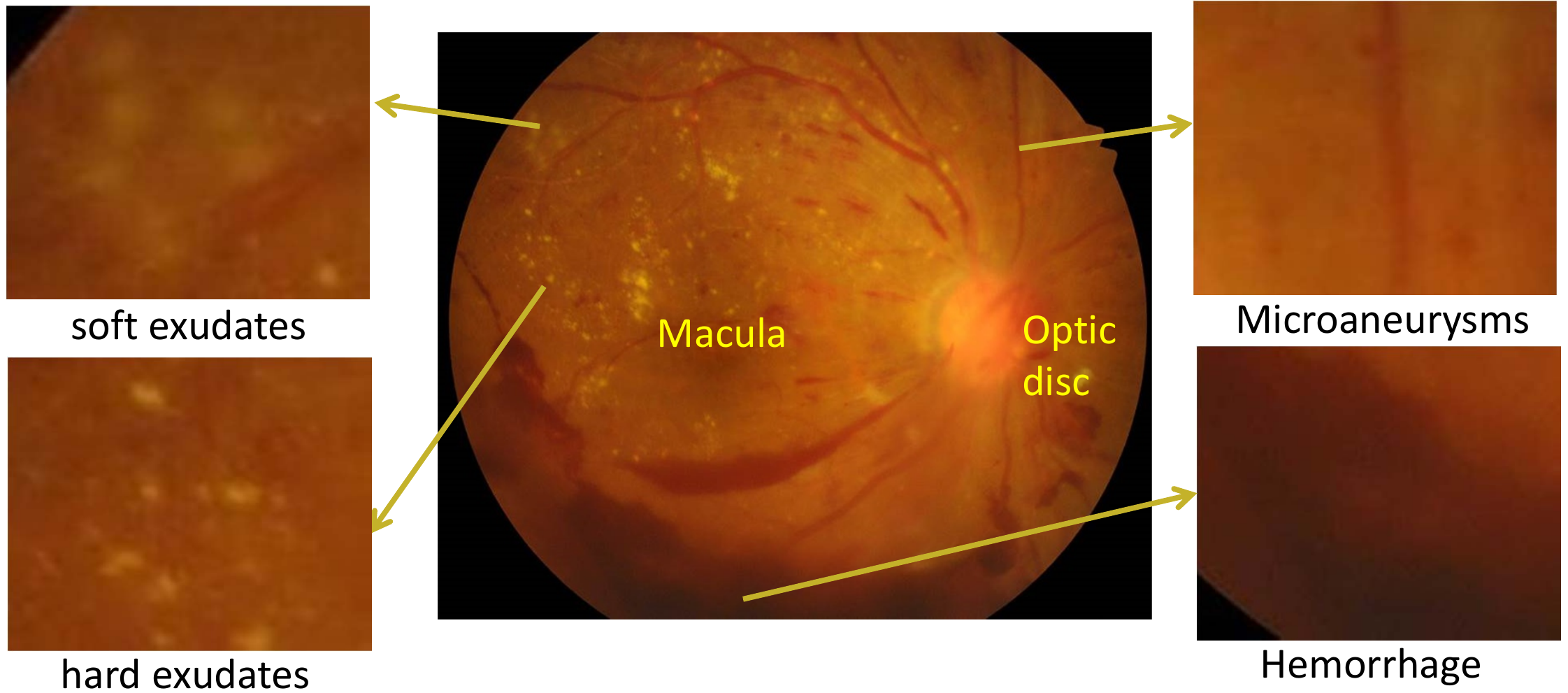}
	\caption{Early pathological signs of DR,~\eg, soft exudates, hard exudates, microaneurysms, and hemorrhage, in a diabetic retinopathy image. Early pathological signs of DME is determined by the shortest distance of macula and hard exudates.} 
	\label{fig:DRsigns}
\end{figure}

\begin{figure*}[t]
	\centering
	\includegraphics[width=1.0\textwidth]{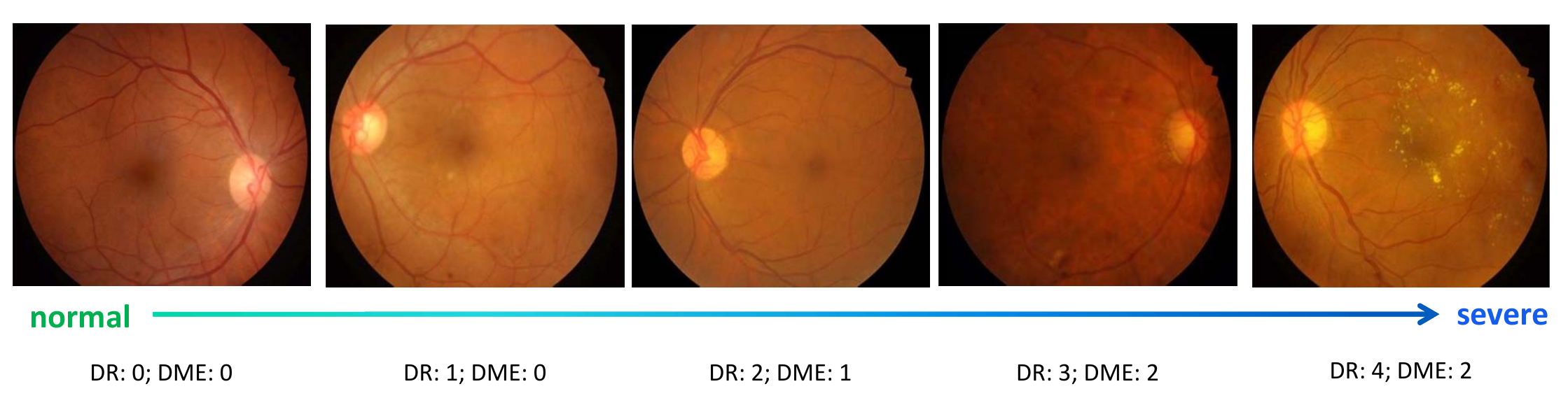}
	\caption{Examples of fundus images with different pathological severity of DR and DME. } 
	\label{fig:examples}
\end{figure*}

Diabetic Retinopathy (DR) is a consequence of microvascular retinal changes triggered by diabetes. It is the most common leading cause of blindness and visual disability in the working-age population worldwide~\cite{cho2018idf}.
Structures such as microaneurysms, hemorrhages, hard exudates, and soft exudates are closely associated with DR and the presence of each of the aforementioned anomaly determines the grade of DR in the patient, as shown in Figure~\ref{fig:DRsigns}.
Diabetic Macular Edema (DME) is a complication associated with DR, which is normally due to the accumulation of fluid leaks from blood vessels in the macula region or retinal thickening that occurs at any stage of DR~\cite{das2015diabetic}. 
\revise{The grading of the severity of DME is based on the shortest distances of the hard exudates to the macula. The closer the exudate is to the macular, the more the risk increases; see examples in Figure~\ref{fig:examples}.}
The most effective treatment for DR and DME is at their early stage, for example, by laser photocoagulation. 
Therefore, in clinical practice, it is important to classify and stage the severity of DR and DME, so that DR/DME patients can receive tailored treatment at the early stage, which typically depends on the grading.



Convolutional neural networks (CNNs) have been proven to be a powerful tool to learn features for DR~\cite{islam2018deep,kang2018multi,krause2018grader} and DME~\cite{ren2018diabetic,syed2018fundus} grading.
For example, Islam~\etal~\cite{islam2018deep} developed a network to detect early-stage and severity grades of DR with heavy data augmentation.
Zhou~\etal~\cite{kang2018multi} presented a multi-cell multi-task learning framework for DR grading by adopting the classification and regression losses.
Regarding the DME grading, Ren~\etal~\cite{ren2018diabetic} presented a semi-supervised learning method with vector quantization. 
These methods, however, adopted different deep networks independently for grading each disease,  ignoring the internal relationship between DR and DME, for example, the DME is the complication of DR.

Recently, some works began to explore joint grading of DR and DME~\cite{gulshan2016development,krause2018grader}.
Gulshan~\etal.~\cite{gulshan2016development} employed the Inception-v3 architecture for DR and DME grading, while Krause~\etal.~\cite{krause2018grader} further improved the performance by utilizing the Inception-v4 architecture.
However, these works focused on the network design and simply regarded the joint grading task as a multi-label problem, without considering the implicit relationship between these two diseases. 
In the medical imaging community, some work~\cite{chen2018multi,tan2018deep,liu2018mtmr} employed multi-task learning to explore the relationship between different diseases (tasks).
%
A key factor for the success in multi-task learning is that \emph{the information among different tasks is shared, thereby promoting the performance of each individual task}.

To explore the feature relationship of DR and DME diseases and improve the grading performance for both diseases, it requires \emph{an understanding of each disease, and also the internal relationship between two diseases}.
To this end, we present a novel deep network architecture, called cross-disease attention network (CANet), to selectively leverage the features learned by the deep convolutional neural network, and produce disease-specific (within each disease) and disease-dependent features (between diseases) for joint DR and DME grading.
In particular, we first develop a disease-specific attention module to select features from the extracted feature maps for individual disease (\ie., DR \& DME).
We then present a disease-dependent attention module to explore the internal relationship between two diseases by learning a set of attention weights, such that a larger weight indicates a higher risk of complication (\eg., DME may lead to worsening DR), and vice versa. 
Through the attention mechanism, our network models the implicit relationship between these two diseases, and improves the joint grading performance.

In summary, our contributions are three folds:
\begin{itemize}
\item We present a novel and effective method, named as cross-disease attention network (CANet), to jointly model the relationship between DR and its complication, \ie, DME. 
To the best of our knowledge, this is the first work for joint modeling the disease and its complication for fundus images. 
\item We propose the disease-specific attention module to selectively learn useful features for individual diseases, and also design an effective disease-dependent attention module to capture the internal relationship between two diseases.

\revise{
\item Experiments on the public IDRiD~\cite{porwal2018indian} challenge dataset and the Messidor~\cite{decenciere2014feedback} dataset show that our CANet method outperforms other methods on grading for both diseases, and achieves the best performance on the IDRiD dataset.}


\end{itemize}

\if 1 
Note that there are some works relying early detection of MA and then perform grading. 
Although these methods achieved promising results in disease severity grading, their methods only grade the severity for one disease, and did not consider the multi-complications in one patient. 
Recently, Safwan~\etal.~\cite{safwan2018classification} shows an ensemble method for joint grading of DR and DME from retinal images. However, there method did not consider the relationship between these two diseases.
Intuitively, we can design four deep learning models (each of which is responsible for the classification of an individual characteristic) and train them separately, which is called single-task learning. Indeed, we previously designed
DeepSeeNet, a deep learning model for the classification of AMD (at the patient level) that uses single-task learning22. However, these four variables are related; training separately may cause the model not to benefit from shared
One intuitive approach is jointly trained the DR and DME classification with the same feature extraction network and two classification loss function, targeting at.

However,

Another simliar work explore the multi-task learning for brain.\cite{namburete2018fully}
The proposed network, which we call the Multi-Task Attention Network
(MTAN) (see Fig. 1), is composed of a single shared network, which learns a
global feature pool containing features across all tasks. 
In this paper, we propose the cross-task attention network, which learns both task-shared and task-specific features.
For each task, rather
than learning directly from the shared feature pool, a soft attention mask is applied at each convolution block in the shared network. In this way, each attention mask automatically determines the importance of the shared features for the respective task, allowing learning of both task-shared and task-specific features in
a self-supervised, end-to-end manner.
\fi

\if 1 
In a computer aided diagnostic system for DME, automated detection of macula and exudates is a
vital task [2–4]. Recently, many kernel based methods with faster optimization speed or stronger generalization performance have been proposed and investigated with theoretic analysis and experimental
evaluation [5,6]. For example, SVM is used to grade the diabetic maculopathy in [2]. An improved fuzzy
C-means along with SVM are utilized for the detection of exudates [7]. Moreover, neural network-based
approaches are proposed to automatically detect exudates in retina images [8]. Although a number of
methods have been proposed for the macula and exudate detection, there are still some unsolved issues.
The major issue is the existence of false positive regions for exudates and macula. Most of the exu-

It consists of three main processing stages: 1) macula localization, containing
several main retina tissue detection, i.e. vessel, optic disc and macula; 2) exudate detection, containing
three steps of exudate candidates’ identification, feature extraction and suspicious candidates classification; 3) grading of diabetic macular edema, according to macular coordinates and location of exudates
to classify the input image into one of the status: normal, Non-CSME and CSME

Tan \etal~\cite{tan2018deep} propose to address the

The risk of having DME is classified into no risk and two probable risks.
International Council of Ophthalmology (ICO) report [5] indicate that 1 out of 3 individuals affected
with diabetes had some form of DR and also highlighted that 1 in 10 had vision-threatening DR.
Currently the world faces a challenge of epidemic disease called Diabetic Macular Edema (DME). 
\\
According to estimations of the World Health Organization, the total numbers
of people in the world with DME will increase to 360 million
by the year 2030 [1]. Diabetes with extended suffering may
lead to diabetic retinopathy (DR) which is consists on the
destruction of blood vessels of the retina. DME is a problem
of DR and is usually one of the main causes of blindness
and visual loss [2], [3], [4]. DME is normally detected and
diagnosed when the fluid leaks from blood vessels inside the
region of macula. The leak is produced by the breakdown
of endothelial tight junctions in blood vessels or microaneurysms

Along with the findings defined in sub-challenge – 1, there may be a presence of venous beading, retinal neovascularization which can be utilized to classify DR retinopathy in one of the two phases known as non-proliferative diabetic retinopathy (NPDR) and proliferative diabetic retinopathy (PDR) as shown in Figure 1a and 1b. DME is a complication associated with DR in which retinal thickening or accumulation of fluid can occur at any stage of DR. The risk of having DME is classified into no risk and two probable risks (illustrated in Figure 1c and 1d respectively) based on the location of hard exudates as mentioned in Table 2. The determination of DR and DME severity based on criteria given in Table 1 and Table 2. It is essential to decide the need for treatment and follow-up recommendations.

Diabetic Retinopathy (DR) and Diabetic Macular
Edema (DME) are the most common eye diseases leading
to eternal vision loss. Therefore early detection is the main
concern to prevent untoward effects by developing
methods for fast detection. In this work, the problem of
automatic classification of Spectral-Domain Optical
Coherence Tomography SD-OCT[1] images have been
addressed for identification of patients with DR, DME
versus normal subjects
Automatic grading of images for DR and DME based on the criteria given above. For this sub-challenge participant will have to submit results for grading of DR and DME both.
Diabetic Retinopathy (DR), one of the leading causes of
blindness in humans, is a consequence of rupture of blood
vessels in the eye and thereby leading to the discharge of
blood and fluid to the surrounding tissues [1]. Structures
such as microaneurysms, hemorrhages and hard exudates
are closely associated with DR and the presence of each 
aforementioned anomaly determines the grade of DR in the
patient. Diabetic Macular Edema (DME) is a condition that
could occur at any stage of DR and is characterized by the appearance of exudates close to the macula or retinal thickening thus affects the central vision of the patient [2]. The treatment administered to subjects with DR or DME is dependent
on the grade of each anomaly and thus classifying the degree
of severity of DR and DME is of utmost importance.
For a variety of classification and pattern recognition
based tasks, Convolutional Neural Networks (CNNs) have
outperformed the traditional machine learning approaches
[3]. The superior performance of the CNNs comes at the cost
of requiring millions of high quality labeled data for training
the network. The presence of huge amount of labeled data
in the domain of medical image analysis is extremely rare.

Test data is the fast adaptationDiabetic Retinopathy (DR) is the consequence of microvascular retinal changes triggered by diabetes and it is the most common leading cause of avoidable blindness in the working-age population in the world.
Structures such as microaneurysms, hemorrhages and hard exudates are closely associated with DR and the presence of each of aforementioned anomaly determines the grade of DR in the patient.
\fi  

%% file: related.tex
\section{Related Work}
\subsection{Diabetic Retinopathy Grading}
Early works on automatic diabetic retinopathy grading were based on the hand-crafted features to measure the blood vessels and the optic disc, and on counting the presence of abnormalities such as microaneurysms,
soft exudates, hemorrhages, and hard exudates,~\etc. 
Then the grading was conducted using these extracted features by different machine learning methods~\cite{silberman2010case,sopharak2009automatic,roychowdhury2013dream,acharya2009computer,akram2014detection,antal2014ensemble,seoud2015red,kumar2017kernel},~\eg., support vector machines (SVM) and k-nearest neighbor (kNN) and Gaussian mixture model. 

In the last few years, deep learning algorithms have become popular for DR grading~\cite{pratt2016convolutional,li2017convolutional,kori2018ensemble,xiao2019major,cao2018efficient,hagos2019transfer,jama2019}. 
There are mainly two categories of deep learning methods for identifying DR severity.
The first category is to use location information of tiny lesions, \eg, microaneurysms, hemorrhage, to determine DR grading performance.
Van Grinsven~\etal~\cite{van2016fast} sped up model training by dynamically selecting misclassified negative samples for hemorrhage detection.
Dai~\etal~\cite{dai2017retinal} proposed a multi-modal framework by utilizing both expert knowledges from text reports and color fundus images for microaneurysms detection.  
Yang~\etal~\cite{yang2017lesion} designed a two-stage framework for both lesion detection and DR grading by using the annotations of locations including microaneurysms, hemorrhage, and exudates.
Lin~\etal~\cite{lin2018framework} developed a new framework, where it first extracted lesion information and then fused it with the original image for DR grading.
Zhou~\etal~\cite{zhou2019collaborative} proposed a collaborative learning method
for both lesion segmentation and DR grading using pixel-level and image-level supervisions simultaneously.

The second category uses image-level supervision to train a classification model to distinguish DR grades directly~\cite{gulshan2016development,gargeya2017automated,wang2017zoom}. Gulshan~\etal~\cite{gulshan2016development} proposed an inception-V3 network for DR grading. 
Gargeya~\etal~\cite{gargeya2017automated} designed a CNN-based model for DR severity measurements.
Wang~\etal~\cite{wang2017zoom} used attention maps to highlight the suspicious regions, and predicted the disease level accurately based on the whole image as well as the high-resolution suspicious patches. 
It is expensive to annotate the labels on the medical images in a pixel-wise manner, hence, we follow the second category to conduct disease grading with only image-level supervision. 


\begin{figure*}[!t]
	\centering
	\includegraphics[width=1.0\textwidth]{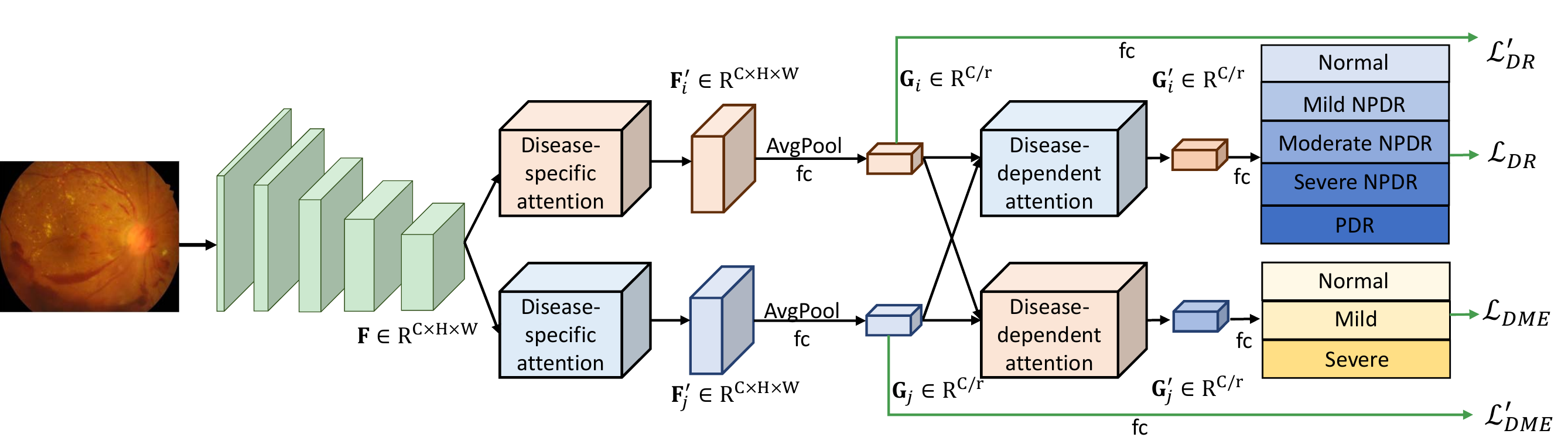}
	\caption{\revise{The schematic illustration of the overall cross-disease attention network (CANet). $\ve{G}_i$ and $\ve{G}_j$ denote the disease-specific features for DR and DME, respectively; $\ve{G}_i^{\prime}$ and $\ve{G}_j^{\prime}$ denote the refined features with disease-dependent information for DR and DME, respectively; $r$ is the ratio to reduce the number of feature channels for saving parameters; fc denotes the fully connected layer; The final loss function is the weighted combination of $\mathcal{L}_{DR}$,  $\mathcal{L}_{DR}^{\prime}$, $\mathcal{L}_{DME}$, and $\mathcal{L}_{DME}^{\prime}$.}}
	\label{fig:framework}
\end{figure*}

\subsection{Diabetic Macular Edema Grading}
Like the DR grading task, grading DME also attracts much attention in the community~\cite{fraz2018computational}.
The assessment of the severity of DME is based on the distances of the exudate to the macula. The closer the exudate is to the macular, the more the risk increases. 
Early works used hand-crafted features to represent the fundus images~\cite{akram2014automated,acharya2017automated}.
For example, Akram~\etal~\cite{akram2014automated} presented a screening system for DME that encompassed exudate detection with respect to their position inside the macular region. The system first extract features for exudate candidate regions, followed by making a representation of those candidate regions. The exact boundaries were determined using a hybrid of GMM model.
However, the capacity of the hand-crafted features is limited. 
CNN based methods~\cite{ren2018diabetic,syed2018fundus} have dramatically improved the performance of DME grading. 
For example, Ren~\etal~\cite{ren2018diabetic} proposed a semi-supervised graph-based learning method to grade the severity of DME.
Syed~\etal~\cite{syed2018fundus} used knowledge of location information of exudates and maculae to measure the severity of DME.  
However, all of these work utilize the location information of exudate regions for disease grading. Such annotations (both lesions masks and grading labels) are difficult to obtain, in this work, we grade DME with only image-level supervision. 
Under the image-level supervision,Al-Bander~\etal~\cite{al2016diabetic} proposed a CNN-based method based on foveae and exudates location for DME screening.
\revise{However, their method classifies the DME into two classes, which is simpler than ours. }

\subsection{Multi-task Learning in Medical Imaging Domain}
Since jointly grading DR and DME diseases is related to the multi-task learning, we also review related works in medical imaging domain~\cite{moeskops2016deep,chen2018multi,tan2018deep,liu2018mtmr,xue2018full} and most of them are designed for image classification or regression tasks.
%
For example, Chen~\etal~\cite{chen2018multi} trained a classification network for four tasks on the Age-related Macular Degeneration disease grading by using the CNN layers to capture common features then fully connected layers to learn the features for individual tasks.
Liu~\etal~\cite{liu2018mtmr} employed a margin ranking loss to jointly train the deep network for both lung nodule classification and attribute score regression tasks.
Similarly, Tan~\etal~\cite{tan2018deep} used the multi-level shared features and designed individual decoders to jointly learn the organ probability map and regressing boundary distance map.
In contrast to these works that jointly do  classification and regression tasks, we design a novel deep network architecture to {\em explore the relationship between two diseases}, and improve the overall grading performance for both diseases.

%% file: method.tex
\section{Methodology}
\begin{figure*}[!t]
	\centering
	\includegraphics[width=1.0\textwidth]{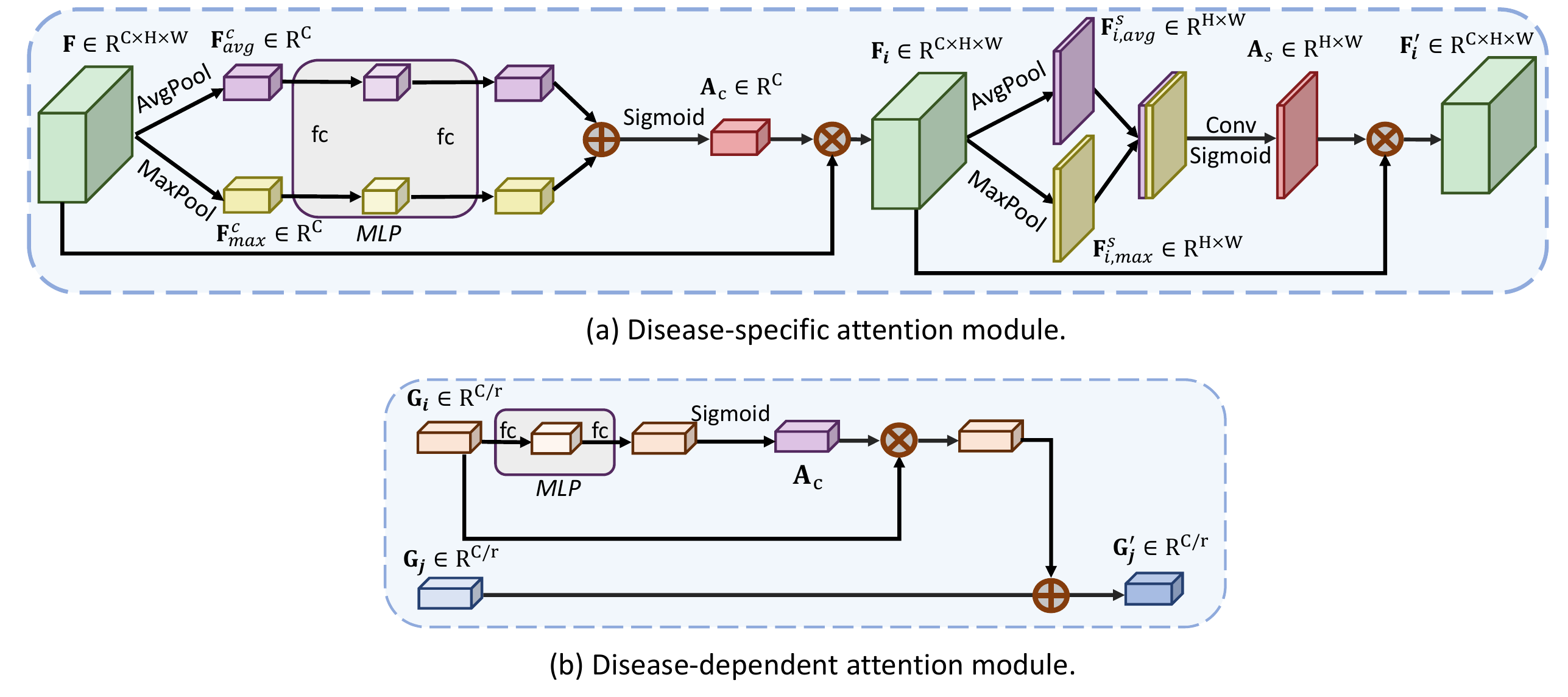}
	\caption{\revise{The architectures of different attention modules. The disease-specific attention module (a) exploits both the inter-channel and inter-spatial relationship of features, while the disease-dependent module (b) explores and aggregates the informative inter-channel features from the other branch; (b) shows an example of disease-dependent attention module used for DME grading; 
	$\ve{A}_c$ denotes the spatial-wise attention 
	map and $\ve{A}_s$ denotes the channel-wise attention map; $r$ is the ratio to reduce the number of feature channels for saving parameters. The actual details of the CANet topology can be found in Table~\ref{table:module_structure}.
} }\label{fig:attention}
\end{figure*}

Figure~\ref{fig:framework} illustrates the overview of our cross-disease attention network (CANet) for joint DR and DME grading, consisting of two disease-specific attention modules (Figure~\ref{fig:attention} (a)) to learn disease-specific features and two disease-dependent attention modules (Figure~\ref{fig:attention} (b)) to explore correlative features between these two diseases.

\subsection{Cross-disease Attention Network}
As shown in Figure~\ref{fig:framework}, our cross-disease attention network takes a fundus image as the input and outputs the grading scores for both DR and DME diseases in an end-to-end manner. 
First, we adopt a convolutional neural network, \ie, ResNet50~\cite{he2016deep} to produce a set of feature maps with different resolutions.
\revise{Then, we take the feature maps  ${\bf F} \in \mathbb{R}^{C \times H \times W}$ with the smallest resolution and highly-semantic information (the deepest convolutional layer in ResNet50) as the inputs for the following two disease-specific attention modules, which learn the disease-specific features ${\bf F}_i^{\prime} \in \mathbb{R}^{C \times H \times W}$ and ${\bf F}_j^{\prime} \in \mathbb{R}^{C \times H \times W}$ to understand each individual disease.
Note that the feature is the one before the AvgPool and fully connected layer of original ResNet. It contains high-level semantic information for DR and DME.	
}
Afterwards, we propose disease-dependent attention modules to explore the internal relationship between the two correlative diseases and produce the disease-dependent features for DR and DME, respectively.  
Finally, we predict the grading scores for DR and DME based on the learned disease-dependent features.

In the following subsections, we will first elaborate the disease-specific attention module and disease-dependent attention module in details, and then present the training and testing strategies of our network for DR and DME grading.

\if 
Note that the features of different dilated convolutional layers at the DDFP contain complementary information for saliency detection due to different dilation rates. 
The dilated convolutional layers with large dilated rates capture global saliency context information (encoding the cues of whole salient objects) but lack of saliency details due to its large receptive field, while the small dilated rates in the dilated convolutional layers help to discover the local saliency context information, which is usually corresponding to the fine saliency detail information but neglects the global semantic information of salient objects. 
Motivated by this observation, we develop an attention block (see Figure 3) to further
refine features at each dilated convolutional layer of the DDFP by fully exploiting the complementary information encoded at different dilated convolutional layers in DDFP. 
\fi
 
\subsection{Disease-specific Attention Module}
Each disease has its specific characteristics, \ie, DR is graded by the presence of soft exudates, hard exudates, hemorrhage, and microaneurysms while DME is determined by the shortest distance between the macula and hard exudates~\cite{xiao2019major}. 
However, the feature maps ${\bf F} \in \mathbb{R}^{C \times H \times W}$ extracted by the convolutional neural network only contain the high-level representations of the input image and it is difficult to capture the specific characteristics for each disease. 
In order to learn the representation of each individual disease, we present a novel disease-specific attention module to learn the specific semantic features of DR and DME, receptively.

\revise{
Figure~\ref{fig:attention} (a) illustrates the detailed structure of the proposed disease-specific attention module, which takes the feature maps ${\bf F} \in \mathbb{R}^{C \times H \times W}$ as the input and adopts the channel-wise attention as well as the spatial-wise attention to highlight the inter-channel and inter-spatial relationship of the features related to each disease.}
\revise{
Specifically, we first squeeze the spatial information from the shared feature maps $\bf F$ via spatial-wise average- and max-pooling operations, and obtain two kinds of global spatial features ${\bf F}_{avg}^{c}$ and ${\bf F}_{max}^{c}$. 
Then, we feed them into a shared \textit{MLP} (multi-layer perception) to produce the channel-wise attention maps ${\bf A_c}$. The channel-wise attention maps ${\bf A_c}$ are described in the following:}
\begin{equation}
{\bf A_c} = \sigma [{\bf W}_1ReLU ({\bf W}_0{\bf F}_{avg}^{c}) + {\bf W}_1ReLU ({\bf W}_0{\bf F}_{max}^{c})] \,
\label{eq1}
\end{equation}
where $\sigma$ is a sigmoid function to normalize the attention weights into $[0,1]$, ${\bf W}_0 \in \mathbb{R}^{C/r \times C}$ and ${\bf W}_1 \in \mathbb{R}^{ C \times C / r}$ are the weights of the shared \textit{MLP}, and 
$r$ is the ratio to reduce the number of feature channels for saving the network parameters and we empirically set it as $0.5$.
After obtaining the learned attention weights ${\bf A_c}$, we multiply it with the original  feature maps ${\bf F} \in \mathbb{R}^{C \times H \times W}$ to produce the disease-specific feature maps $\ve{F}_i$:
\begin{equation}
\ve{F}_i = {\bf A_c} \otimes \ve{F} \ ,
\end{equation}
where $\otimes$ denotes the element-wise multiplication, and the attention weights ${\bf A_c}$ are broadcasted along the spatial dimension. 
Hence, we can select the disease-specific features and suppress the features that are irrelevant to the disease along the feature channels.

\begin{table}[!t]
	\begin{center}
		\caption{\revise{The detailed structure of cross-disease attention modules. ``FC'' represents the fully connected layer; ``Conv'' represents the convolution operation; ``ReLU'' and ``Sigmoid'' are the ReLU and Sigmoid non-linear operations, respectively; ``Concat'' represents the concatenation operation. For ``Conv'', we use padding to keep the size of the feature maps. The symbols are defined in Figure 4.}}
		\label{table:module_structure}
		\resizebox{1.0\linewidth}{!}{
			\begin{tabular}{c|c|c|c}
				\toprule[1.5pt]
				&	Input feature &    Type &  Output feature   \\
				\hline
				\multirow{9}{*}{\rotatebox[origin=c]{90}{\parbox[c]{2.2cm}{disease-specific}}}		
				&	$\textbf{F}$ & MaxPool, Flatten   &  $\textbf{F}_{max}^{c} $ \\ 
				
				&	$\textbf{F}$ & AvgPool, Flatten   & $\textbf{F}_{avg}^{c} $ \\ 	
				\cline{2-4} 
				& $\textbf{F}_{avg}^{c}$, $\textbf{F}_{max}^{c}$ & FC 1 ($ 2048 \times 128$), ReLU  & -  
				\\ 
				&	-   &  FC 2 ($ 128 \times 2048$) & -  \\
				
				&	 - &  Sum, Sigmoid    & $\textbf{A}_{c}$  \\	
				\cline{2-4} 
				&	$\textbf{A}_{c} , \textbf{F} $  & Multiplication &   $\textbf{F}_{i}$   
				\\
				
				&		$\textbf{F}_{i}$  &   AvgPool &    $\textbf{F}_{i, avg}^{s}$  \\
				
				& $\textbf{F}_{i}$  & MaxPool  &   $\textbf{F}_{i, max}^{s}$  \\	 
				
				& 	$\textbf{F}_{i, avg}^{s}$, $\textbf{F}_{i, max}^{s}$  & Concat, Conv, Sigmoid  &     $\textbf{A}_{s}$  \\

				& $\textbf{A}_{s}$ ,	$\textbf{F}_{i}$   & Multiplication &   $\textbf{F}_{i}^{\prime}$  \\
				
				\hline

				\multirow{6}{*}{\rotatebox[origin=c]{90}{\parbox[c]{0.4cm}{disease-dependent}}	}

				&  $\textbf{G}_{i}$ &    FC 1 ($1024\times 64$), ReLU  & - \\
				
				& -  &    FC 2 ($64\times 1024$)   & - \\
				
				& -  &   Sigmoid &  $\textbf{A}_c$ \\
				\cline{2-4}
				& $\textbf{A}_c$,  $\textbf{G}_j$  &   Multiplication  &  $\textbf{G}_{j}^{\prime}$ \\
				
				\bottomrule[1.5pt]		
		\end{tabular} }
	\end{center}
\end{table}

\revise{
To further highlight the disease-specific features across the spatial domain, we follow~\cite{woo2018cbam,hu2018squeeze} and adopt another attention model, which aggregates the channel-wise information by applying the max-pooling and avg-pooling operations along the channel dimension and produces the feature maps ${\bf F}_{i,avg}^{s}$ and ${\bf F}_{i,max}^{s}$.
Then, we concatenate these two feature maps together and use another convolutional operation to learn the 2D spatial-wise attention map ${\bf A_s}$:
\begin{equation}
{\bf A_s} = \sigma (Conv([{\bf F}_{i,avg}^{s}; {\bf F}_{i,max}^{s}]) \ ,
\label{eq2}
\end{equation}
where \textit{Conv} is a convolution layer and $\sigma$ denotes the sigmoid function.}
Finally, we obtain the disease-specific features $\ve{F}_i^{\prime}$  ($\ve{F}_j^{\prime}$ for another disease; see Figure~\ref{fig:framework}) by multiplying the learned attention weights ${\bf A_s}$ with the feature maps $\ve{F}_i$ to select the disease-specific features across the spatial dimension:
\begin{align}
\ve{F}_i^{\prime} = {\bf A_s} \otimes \ve{F}_i \ .
\end{align}
Note that the attention weights ${\bf A_s}$ are broadcasted along the channel dimension during the multiplication.
In this way, we can further selectively use the disease-specific features by enhancing the disease-relevant features and suppressing the disease-irrelevant features across the spatial domain.

\revise{We show the detailed structure of the disease-specific attention module in Table~\ref{table:module_structure}. 
The input and output channel number of FC 1 and FC 2 in disease-specific module are $2048 \times 128$ and $128 \times 2048$, respectively. We use ReLu activation after the first fully connected layer in each attention module. }

\subsection{Disease-dependent Attention Module}
As the statistics of the grading labels shown in Table~\ref{tab:messidor_dataset} and Table~\ref{tab:idrid_dataset}, DR and DME have the internal relationship.
\revise{On the one hand, the more exudates are, the greater risk of the macula may have,~\ie, severer of DR may lead to severer DME.}
On the other hand, the closer of exudates to the macula, the more risk of presences of pathological DR signs,~\ie, worser of DME may lead to worser DR. 
Motivated by this observation, we present the disease-dependent attention module (see Figure~\ref{fig:attention} (b)) to capture the internal relationship between these two diseases.
\begin{table}[h]
	\centering
	\caption{\revise{The statistics of the labels in the Messidor dataset. The first number is the counts of labels and the second one is the relative value.}}{
{\begin{tabular}{c|c|c|c|c}
		\toprule[1.5pt] 
		\diagbox{DME}{DR} & 0	& 1 & 2 &  3  \tabularnewline
		\hline 
		0	& 546, 45.5\%	&  142, 11.8\% & 182, 15.2\% & 104, 8.7\% \tabularnewline 
		1	& 0, 0.0\%	& 5, 0.4\%   & 28, 2.3\% &   42, 3.5\% \tabularnewline 
		2	& 0, 0.0\% & 6, 0.5\%  & 37, 3.1\% & 108, 9.0\% \tabularnewline 		
		\bottomrule[1.5pt]	 
	\end{tabular}}}
	\label{tab:messidor_dataset}
\end{table}	
\begin{table}[h]
	\centering
	\caption{\revise{The statistics of the labels in the IDRiD dataset. The first number is the counts of labels and the second one is the relative value. }}{
\resizebox{1.0\linewidth}{!}{	{\begin{tabular}{c|c|c|c|c|c}
			\toprule[1.5pt] 
			\diagbox{DME}{DR} & 0	& 1 & 2 & 3 & 4  \tabularnewline
			\hline 
			0	& 134, 26.0\%	&  18, 3.5\% &  36, 7.0\%  &  5, 1.0\% &  4, 0.8\% \tabularnewline 
			1	& 0, 0.0\%	& 0, 0.0\%  & 24, 4.6\%  & 4, 0.8\%  & 2, 0.4\%  \tabularnewline 
			2	& 0, 0.0\% & 0, 0.0\%  & 140, 27.1\% & 116, 22.4\%  &  33, 6.4\% \tabularnewline 		
			\bottomrule[1.5pt]	 
	\end{tabular}}}}
	\label{tab:idrid_dataset}
\end{table}	

As shown in Figure~\ref{fig:framework}, this model takes the disease-specific features of both DR and DME diseases as the inputs, \ie, $\ve{G}_i$ and $\ve{G}_j$, which are obtained by adopting the average pooling and fully connection operations on $\ve{F}_i^{\prime}$ and $\ve{F}_j^{\prime}$, and then it learns to produce the disease-dependent features for DR or DME, respectively. 
Figure~\ref{fig:attention} (b) illustrates the detailed structures of the proposed disease-dependent attention module used for DME grading, which has the similar structures to the attention model used for DR grading.


Specifically, given the feature maps $\ve{G}_i$ of DR disease, we first employ a \textit{MLP} and a sigmoid function to learn a set of attention weights ${\bf A}_{DR}$, and then multiply these weights with the input feature maps $\ve{G}_i$ to select the useful features, which helps to identify the DME disease.
After that, we add the selected feature maps with the specific features of DME disease $\ve{G}_j$ in an element-wise manner ($\oplus$) to generate the disease-dependent features of DME $\ve{G}'_j$:
%
\begin{align}
	\label{eq3}
	{\bf A}_{DR} &= \sigma [{\bf W}_1^{DR}ReLU ({\bf W}_0^{DR}(\ve{G}_i))] \ , 
\\ 
	\ve{G}'_j &=\ve{G}_j \oplus {\bf A}_{DR} \otimes \ve{G}_i \ . 
\end{align}

Hence, the network is able to capture the correlation between the DR and DME diseases and improves the overall grading performance for both DR and DME diseases.
\revise{The detailed structure of disease-dependent attention module is shown in Table~\ref{table:module_structure}. 
The input and output channel number of FC 1 and FC 2 are  $1024 \times 64$ and $64 \times 1024$, respectively. We use ReLu activation after the first fully connected layer in the attention module.}



\revise{

\subsection{Network Architecture}}
We adopted ResNet50 as the backbone network to extract features, followed by a dropout layer with the drop rate of $0.3$, and employed two disease-specific attention modules to learn disease-specific features. 
We employed two loss functions,~\ie, $\mathcal{L}_{DR}^{\prime}$ and $\mathcal{L}_{DME}^{\prime}$, to learn disease-specific features, and another two loss functions,~\ie, $\mathcal{L}_{DR}$ and $\mathcal{L}_{DME}$, for the final DR and DME grading:
\begin{align}
\label{eq_loss}
\mathcal{L} & = \mathcal{L}_{DR}  + \mathcal{L}_{DME}  + \lambda (\mathcal{L}_{DR}^{\prime}  + \mathcal{L}_{DME}^{\prime}),
\end{align}
where $\mathcal{L}_{DR}^{\prime}$ and $\mathcal{L}_{DME}^{\prime}$ denote the cross-entropy loss for DR-specific and DME-specific feature learning, respectively; $\mathcal{L}_{DR}$ and $\mathcal{L}_{DME}$ denotes the loss function for the DR and DME grading.
\revise{$\mathcal{L}_{DR}$ is a binary cross-entropy loss on the Messidor dataset and a 5-class cross-entropy loss on the IDRiD dataset. $\mathcal{L}_{DME}$ is a three-class cross-entropy loss on both Messidor and IDRiD dataset. 
\revise{
	\begin{equation}
	\label{eq: weighted loss}
	\mathcal{L}(y,\hat{y}) = -\frac{1}{N}\sum_{i=1}^{N}\sum_{c=1}^{M} y_i^c \log \hat{y_i}^c
	\end{equation}
where $\hat{y_i}^c$ denotes the probability of voxel $i$ belongs to class/grade $c$, and $y_i^c$ indicates the ground truth label for retinal image $i$. 
M is three for DME grading and two or five for DR grading (two in the Messidor dataset and five in the IDRiD dataset). 
} 	
Taking DR as an example, we directly apply a fully connected layer on DR-specific features $\ve{G}_i$ (batch size $\times$ 1024) for classification. The kernel size of fully connected layer is 1024 $\times$ 2 for Messidor dataset and 1024 $\times$ 5 for IDRiD dataset.}
$\lambda$ is the weight in the loss function.
When $\lambda$ = 0.0, the network is optimized by the refined DR and DME features that include both disease-specific and disease-dependent information.  
As $\lambda$ increasing, the framework gives more importance to the disease-specific feature learning. We analyze the effects of different $\lambda$ in the experiment part, and we empirically set $\lambda$ as $0.25$.

\revise{
\subsection{Training and Testing Strategies}
We normalized the training images and resize images to $350 \times 350 $ resolution. 
For data augmentation, we randomly scaled and cropped the images into the patches with a size of $224 \times 224$. Random horizontal flip and vertical flip were also used to augment the training data.
We optimized the network with Adam optimizer~\cite{kingma2014adam}. The initial learning rate was 0.0003 and we decayed the learning rate with a cosine annealing for each batch~\cite{loshchilov2016sgdr}. 
We trained the network for 1000 epochs and the batch size is 40.
During the training process, we feed the samples of DR and DMR in a random order.
The whole framework was built on  PyTorch~\cite{paszke2017automatic} with Titan Xp GPU. The network has 29 M trainable parameters.
The training time of the network was five hours and the inference time was 0.02 seconds per image. 

To test the grading result, we only used the prediction score after the refined DR and DME features, which include the disease-dependent information.
We selected the class with the maximum prediction value in DR and DME, respectively.
During inference, we did not use any post-processing operations and model ensemble techniques.  
}

%% file: experiments.tex
\begin{table*}[!t]
	\centering
	\caption{\revise{Quantitative results on the Messidor dataset. The reported results are the mean values of 10-fold cross validation. Ac, Pre, Rec, F1 denote accuracy, precision, recall, F1-score, respectively (unit: \%)}.
	}{
	{\resizebox{0.98\textwidth}{!}{\begin{tabular}{c|c|c|ccccc|ccccc}
		\toprule[1.5pt]
		\multirow{2}{*}{Methods}  & \revise{\multirow{2}{*}{Parameters}} &\multirow{2}{*}{Joint Ac}	&    \multicolumn{5}{c|}{ DR } & \multicolumn{5}{c}{ DME }  \tabularnewline
		\cline{4-13}
		& 	&  &  AUC & Ac & Pre & Rec & F1 & AUC & Ac & Pre &  Rec & F1      \tabularnewline
		\hline
		Individual training (DR) &  23.52 M & - & 89.5 & 81.0 & 76.2 & 78.9 & 77.3 & -  & - & - & - & -  \tabularnewline
		Individual training (DME) &  23.52 M  &  - & - & - & - & - & - & 89.1 & 86.8 & 62.8 & 65.2 & 61.9   \tabularnewline
		
		Joint training & 23.52 M &  82.0  & 94.2 & 89.1 & 86.5 & 88.2 & 87.2 & 90.5 & 90.4 & 78.7 & 73.9 & 75.3  \tabularnewline
		\revise{Joint training (complex)} & 29.04 M &  82.8  & 95.2 & 91.0 & 90.5 & 88.6 & 89.5 & 90.3 & 91.7 & 73.6 & 71.2 & 71.1  
		\tabularnewline 
		\revisesecond{Joint training (complex\_v2)} & 29.08 M &  82.5  & 95.4 & 90.3 & 87.5 & 89.8 & 88.5 & 89.4 & 91.0 & 78.3 & 72.0 & 73.4  
		
		\tabularnewline
		\hline
		CANet (d-S only) & \multirow{4}{*}{29.03 M} & 84.1  & 95.8 & 91.7 & 91.5 & 88.8 & 89.9 & 90.3 & 91.0 & 79.6  & 71.0 & 72.5  \tabularnewline 	
		CANet (d-S; d-D {\tiny{DR$\Rightarrow$DME}} ) & & 84.5  & 96.0 & 91.7 & 91.5 & 88.7 & 90.0& 89.4 & 91.9 & 82.3 & 72.0& 74.8 \tabularnewline 	
		CANet (d-S; d-D {\tiny{DR$\Leftarrow$DME}} ) &  & 84.9  &  96.3 & 91.9 & 90.8 & 90.1 & 90.3 & 91.0 & 91.9 & 79.6 & 73.2 & 74.6 \tabularnewline 	
		
		\textbf{ CANet ($\lambda$=0.25; final model)} &   & \textbf{85.1} &   96.3 & 92.6 & 90.6 & \textbf{92.0} & \textbf{91.2} & \textbf{92.4} & 91.2 & 76.3 & 70.8 & 72.4 \tabularnewline	
		\hline 
		\hline
		CANet ($\lambda$=0.00)& \multirow{4}{*}{29.03 M} &  84.8  & 96.3 & 92.1 & 92.2 & 88.6 & 90.3 & 91.6 & 91.5 & \textbf{82.4} & 73.7 & 75.3   \tabularnewline
		CANet ($\lambda$=0.50) &  & 84.7 &   96.1 & 91.4 & 89.9 &  89.7 & 89.6 & 92.2 & \textbf{92.0} & 78.6 & \textbf{74.9} & 75.3  \tabularnewline
		CANet ($\lambda$=0.75) &  & 84.8 &  \textbf{96.5} & 92.2 & \textbf{92.3} & 88.8 & 90.3 & 91.8 & \textbf{92.0} & 79.5  & 71.4 & 73.7  \tabularnewline
		CANet ($\lambda$=1.00) &   & 84.9 &   96.3 & \textbf{92.7}  & 91.6 & 90.6 & 91.0 & 90.7 & 91.7 & 77.4 & 69.4 & 71.0  \tabularnewline
		\bottomrule[1.5pt]

%
%

	\end{tabular}}}}
	\label{tab:ablationstudy}
\end{table*}

\section{Experiments and Results}
\subsection{Datasets}
We evaluate the effectiveness of our method by comparing it against existing works on~\textbf{Messidor dataset}~\cite{decenciere2014feedback}\footnote{http://www.adcis.net/en/third-party/messidor/} and \textbf{2018 ISBI IDRiD challenge dataset}~\cite{porwal2018indian}\footnote{https://idrid.grand-challenge.org/Grading/}.
To the best of our knowledge, these two datasets are the only two public datasets with both DR and DME severity grading annotations.

\para{Messidor Dataset.} \ This dataset has 1200 eye fundus color numerical images of the posterior pole acquired from three ophthalmologic departments. For each image in the dataset, its grading annotations of DR and DME are provided by the medical experts to measure the retinopathy grade and risk of macular edema.
Specifically, DR is graded into four classes by the severity scale. \revise{
Given the fact in the DR screening that the difference between normal images and images of stage 1 is the most difficult task for both the CAD systems and clinical experts, S{\'a}nchez~\etal~\cite{sanchez2011evaluation} grouped stages 0 and 1 of the Messidor dataset as referable images and combined stages 2 and 3 as non-referable in their screening work.}
This two-class setting has been widely used in the existing DR screening methods~\cite{vo2016new,wang2017zoom}, so that we conducted binary classification for DR grading in the Messidor dataset.
To fairly compare with previous works~\cite{pires2015beyond,vo2016new,wang2017zoom}, we use 10-fold cross validation on the entire dataset.
DME is annotated based on the shortest distance $d$ between the hard exudates location and the macula.
The severity of DME is graded to 0 (No visible hard exudate), 1 ( $d > $ 1 papilla diameter), 2 ( $d <= $ 1 papilla diameter). 
The statistics of the DR and DME labels in the Messidor dataset is shown in Table~\ref{tab:messidor_dataset}.

\para{IDRiD Dataset.} 
We employed the ISBI 2018 IDRiD sub-challenge 2 dataset.
This dataset includes 516 images with a variety of pathological conditions of DR and DME, consisting of 413 training images and 103 test images. 
In the IDRiD dataset, each image contains both DR and DME severity grading labels.
DR grade is annotated into five classes according to the severity scale, and we perform 5 class classification for DR. 
DME is annotated based on the shortest distance $d$ between the hard exudates location and the macula.
The annotation criteria of DME grading is the same as that in the IDRiD dataset.
The statistics of the labels in the IDRiD dataset is shown in Table~\ref{tab:idrid_dataset}.
The detailed grading criterion for the IDRiD dataset can be found in the provided dataset websites.
Note that we report 10-fold cross validation results for the Messidor dataset and use train \& test sets split by the challenge organizers for the IDRiD dataset.
\subsection{Evaluation Metrics}
To measure the joint grading performance, we employ the IDRiD challenge evaluation metric ``Joint Accuracy" (\textbf{Joint Ac}). 
The definition of \textbf{Joint Ac} is: If the prediction matches both DR and DME ground-truth label, then it is counted as one, else zero. The total number of true instances is divided by a total number of images to get the final result.
We use Joint Ac to select our final model. 
For the Messidor dataset, we also report the accuracy (Ac), AUC, precision (Pre), recall (Rec), F1-score (F1) for each disease. 

For the 2018 ISBI IDRiD dataset, we follow the challenge description and use the challenge evaluation metric (``Joint Ac'') for comparison.


\subsection{Analysis of Network Design}
\subsubsection{Compare with Baselines}
We first compare our method with two baselines, \ie, ``Individual training'' and ``Joint training'' on the Messidor dataset.
``Individual training (DR)'' and ``Individual training (DME)'' indicates that we trained two individual ResNet50 networks for DR and DME grading, respectively.
The ``Joint training'' denotes that we employed a ResNet50 network for shared feature extraction and two individual fully connected layers for DR and DME grading, respectively.

Table~\ref{tab:ablationstudy} reports the 10-fold cross validation results of accuracy, AUC, precision, recall, F1-score for DR and DME respectively, as well as Joint Ac.
It is observed that ``Individual training (DR)'' and ``Individual training (DME)'' achieve 89.5\% AUC and 89.1\% AUC for DR and DME, respectively.
``Joint training'' improves the individual training to 94.2\% AUC and 90.5\% AUC for DR and DME, respectively.
Notably, our method (CANet) with the same backbone (ResNet50) and training strategies  improves the performance over these two baselines, with 96.3\% AUC (DR) and 92.4\% AUC (DME).
The results show that the effectiveness of our method compared with these two baselines.

\revise{
From the listed model parameters in Table~\ref{tab:ablationstudy}, we can see that our method has more parameters (29.03 M), compared with the Joint training (23.52 M). 
To validate the effectiveness of our design under the same model complexity, we increase the parameters of ``joint training'' to 29.04 M by adding several standard components before classification on ``joint training''.  These components include a convolutional layer with kernel size $2048 \times 300 \times 3 \times 3$, batch normalization layer and ReLU activation.
\revisesecond{We also implemented another complex joint training baseline, \ie, ``Joint training (complex\_v2)'' in Table~\ref{tab:ablationstudy}. This architecture is implemented by adding three convolutional layers on ``Joint training'' baseline. Specifically, the convolutional layer has the filter shapes of $2048 \times 660 \times 1 \times 1$, $660 \times 512 \times 3\times 3$, and $512\times 256 \times 3\times 3$, respectively. Each convolutional layer is followed by a BN and a ReLU activation. 
Such complex baselines achieve 82.8\% and 82.5\% on joint Ac, respectively. However, with the same level of network parameters, our method (85.1\%) still achieves the best performance, showing the effectiveness of the attention modules.}
}

\begin{table*}[t]
	\centering
	\caption{Comparison with other multi-task learning methods on the Messidor  dataset. The reported results are the mean of 10-fold cross validation (unit: \%).
	}{
	{\resizebox{0.9\textwidth}{!}{\begin{tabular}{c|c|ccccc|ccccc}
		\toprule[1.5pt]
		\multirow{2}{*}{Methods}   &\multirow{2}{*}{Joint Ac}	&    \multicolumn{5}{c}{ DR } & \multicolumn{5}{c}{ DME }  \tabularnewline
		\cline{3-12}
		& 	 &  AUC & Ac & Pre & Rec & F1 & AUC & Ac & Pre &  Rec & F1      \tabularnewline
		\hline

		
		Multi-task net~\cite{chen2018multi} & 82.4 & 94.8 & 89.9 & 89.7  & 85.7 & 87.5 & 90.5  & 90.5 & 79.1 & 70.8 & 72.2  \tabularnewline
		MTMR-Net~\cite{liu2018mtmr} & 83.1 & 94.9 & 90.3 & 90.0 & 86.7 &  88.1 & 90.6 & 90.4 &\textbf{79.8}& \textbf{73.2} &  \textbf{75.4}  \tabularnewline
		\textbf{CANet (ours)}  &  \textbf{85.1} &  \textbf{96.3} & \textbf{92.6} & \textbf{90.6} & \textbf{92.0} & \textbf{91.2} & \textbf{92.4} & \textbf{91.2} & 76.3 & 70.8 & 72.4  \tabularnewline
		
		\bottomrule[1.5pt]	
	\end{tabular}}}}
	\label{tab:multitask}
\end{table*}

\subsubsection{Analyze the Attention Module}
We analyze the effects of disease-specific and disease-dependent attention modules.
The comparisons are conducted with the same network backbone (ResNet50) and training strategies.  The results are reported in Table~\ref{tab:ablationstudy} by the 10-fold cross validation on the Messidor dataset.
Compared with the ``Joint training'', adding the disease-specific attention module, \ie, ``CANet (d-S only)'' enhances the Joint Ac from 82.0\% to 84.1\%.
The accuracy of DR and DME are also improved from 89.1\% to 91.7\% (DR) and from 90.4\% to 91.0\% (DME), respectively.
These comparisons demonstrate that disease-specific attention module explores more discriminative features for specific disease grading.

Then, we analyze the importance of the disease-dependent attention module for DME, \ie, 
``CANet (d-S; d-D {\tiny{DR$\Rightarrow$DME}})".
This experiment indicates that the correlative feature learned on DR is incorporated to DME branch, and vice versa.
It is observed that DR$\Rightarrow$DME improves the Joint Ac result to 84.5\%, and DME grading results are enhanced on most evaluation metrics.
Furthermore, we also analyze the importance of the disease-dependent attention module for DR, \ie, ``CANet (d-S; d-D {\tiny{DR$\Leftarrow$DME}})''. 
With this dependent attention branch, the joint accuracy is boosted to 84.9\%, and DR grading results are also increased on most evaluation metrics. 
When we incorporate the disease-dependent attention module into both branches, our method ``CANet ($\lambda$=0.25; final model)'' achieves the highest results, with joint Ac of 85.1\%.
These results validate that the disease-specific and disease-dependent attention module are both effective to utilize the disease-specific and disease-dependent information for better joint grading.

\subsubsection{Analyze the Weight $\lambda$ in the Loss Function}
We analyze the effect of the weight $\lambda$ in our method.
The bottom part of Table~\ref{tab:ablationstudy} shows the results with different weights in the loss function. When $\lambda$ = 0.00, that whole framework is trained with the final refined DR and DME features that include the both specific and dependent information.
When $\lambda$ increases, the network is trained with the additional supervision for disease-specific attention.
As shown in the Table~\ref{tab:ablationstudy}, the variance of results with different $\lambda$ is little, which indicates that our method is not very sensitive to the weight in the loss function.
Our method reaches the best ``Joint Ac'' result (85.1\%), when $\lambda$ = 0.25. Therefore, we choose this model as our final model. 

\subsubsection{Analysis on Architectures}
\revise{To analyze the effectiveness of backbone models, we perform experiments on ``Joint training'' to select the proper backbone architecture. 
The ``Joint training'' denotes that we employed
a backbone network for shared feature extraction and two individual fully connected layers for DR and DME grading, respectively.
We implemented with ResNet50~\cite{he2016deep}, ResNet34~\cite{he2016deep}, and DenseNet161~\cite{huang2017densely} and the results is showed in Table~\ref{tab:achitect}.
We can see that ResNet50 achieves better results and finally we use ResNet50 as the backbone model.

\begin{table}[h]
	\centering
	\caption{\revise{Results of different backbone architectures on the Messidor dataset (unit: \%).}}
	\resizebox{0.5\linewidth}{!}{\begin{tabular}{ccc}
			\toprule[1.5pt]
			Methods  & Joint Ac & \tabularnewline \hline 
			ResNet50  & 82.0 & \tabularnewline 
			ResNet34 & 81.4 & \tabularnewline 
			DenseNet161 & 78.9 & \tabularnewline 
			\bottomrule[1.5pt]
		\end{tabular}}
		\label{tab:achitect}
\end{table}
}	

\subsection{Compare with other Multi-task Learning Methods}

To the best of our knowledge, there is no previous work for joint DR and DME grading. To show the effectiveness of our method for joint grading, we compare our method with two recent multi-task learning methods in the medical imaging community.
Chen~\etal~\cite{chen2018multi} designed a method for the Age-related Macular Degeneration disease grading, while Liu~\etal~\cite{liu2018mtmr} proposed a network for both lung nodule classification and attribute score regression tasks. 
Since these works are not tailored for DR and DME grading, we did not directly use their methods for joint DR and DME grading. Instead, we adapted their key ideas to our task with the same network backbone and training strategies for fair comparison.
\revise{For~\cite{chen2018multi}, after the ResNet50 feature extractor, we use the average pooling operation. Then, we use another one fully connected layer to reduce the channel number to 1024, followed by three fully connected layers (channel number: 1024, 256, 128) for DR and DME grading, respectively. The dropout layer is also employed.
For~\cite{liu2018mtmr}, we use a fully connected layer with channel size 256 to concatenate the information from one task to another task, then two individual fully connected layers are employed for final DR and DME grading, respectively.}

We report the performance of these two methods in Table~\ref{tab:multitask}.
It is observed that our method clearly outperforms these multi-task learning based methods on the Joint Accuracy metric.
Compared with~\cite{liu2018mtmr}, our method achieves 1.4\% (AUC) and 2.3\% (Ac) improvement for DR; 1.8\% (AUC) and 0.8\% (Ac) improvement for DME.
These results show the superiority of our framework for joint DR and DME grading.

\begin{table}[h]
\centering
\caption{\revise{Results of different methods on the Messidor dataset. Our result is under 10-fold cross validation. Other results are copied from original papers. ``-'' indicates no reported result.} }{
{\resizebox{0.85\linewidth}{!}{
		\begin{tabular}{c|cc|cc}
			\toprule[1.5pt]
			\multirow{2}{*}{Methods}  
			&    \multicolumn{2}{c}{ DR } & \multicolumn{2}{c}{ DME }   \tabularnewline
			\cline{2-5}
			&  AUC & Ac & AUC & Ac    \tabularnewline
			\hline 	
			
			Lesion-based~\cite{pires2015beyond}  & 76.0 & -&  - &  -  \tabularnewline
			
			Fisher Vector~\cite{pires2015beyond}  & 86.3 & - & - & -  \tabularnewline
			
			VNXK/LGI~\cite{vo2016new}  & 88.7 & 89.3 & - & - \tabularnewline
			
			CKML Net/LGI~\cite{vo2016new} & 89.1 & 89.7 & - & -     \tabularnewline
			
			Comprehensive CAD~\cite{sanchez2011evaluation} & 91.0 & - & - & - \tabularnewline
			
			DSF-RFcara~\cite{seoud2015red}  & 91.6 & - & - & -    \tabularnewline
			\revise{Clinical B~\cite{sanchez2011evaluation}  } & 92.0 & - & - & -   \tabularnewline
			\revise{Clinical A~\cite{sanchez2011evaluation}} & 94.0 & - & - & -   \tabularnewline
			
			Zoom-in-net~\cite{wang2017zoom}$~\dagger$  & 95.7 & 91.1  & - &  - \tabularnewline
			DME classifier~\cite{al2016diabetic}  & - & - & - &  88.8  \tabularnewline
			
			\textbf{CANet (ours)}   &  \textbf{96.3} & \textbf{92.6}  &  \textbf{92.4} & \textbf{91.2}  \tabularnewline
			\bottomrule[1.5pt]	
			
		\end{tabular}}
		\begin{tablenotes}
			\centering
			\item ``$\dagger$'': denotes using additional  dataset EyePACS~\cite{cuadros2009eyepacs} as the pretrain.
		\end{tablenotes}
		\label{tab:messidor}}}
\end{table}	
\subsection{Comparisons on the Messidor Dataset}
We also compare our method with other DR grading models and DME grading models reported on the Messidor dataset in Table~\ref{tab:messidor}.
As described in section II, there are two main branches for DR grading: employing both image-level and lesion location information as the supervision~\cite{lin2018framework,zhou2019collaborative,ramachandran2018diabetic,sahlsten2019deep}, and employing only image-level supervision~\cite{sanchez2011evaluation,vo2016new,wang2017zoom}. As for DME grading, some works~\cite{acharya2017automated,ren2018diabetic,syed2018fundus} utilized macular or lesion location information features to help the grading of DME. For fair comparison, we only compare with those methods with only image-level supervision.

\revise{For DR grading models, the combined kernels with multiple losses network (CKML)~\cite{vo2016new} and VGGNet with extra kernels (VNXK)~\cite{vo2016new} aims to employ multiple filter sizes to learn fine-grained discriminant features.
Moreover, clinical experts~\cite{sanchez2011evaluation} were also invited to grade on the Messidor dataset.
It is worth mentioning that our method outperforms the clinical experts by 2.3\% abd 4.3\% on the AUC metric.
Note that the clinical experts are provided by specific expert in~\cite{sanchez2011evaluation}. 
Recently, Wang~\etal~\cite{wang2017zoom} proposed the gated attention model and combined three sub-networks to classify the holistic image, high-resolution crops and gated regions.}
It is worth noticing that they first pretrain their model on EyePACS dataset~\cite{cuadros2009eyepacs} and then fine tune on the Messidor dataset, while we only use the Messidar dataset to train our model.  
Our method with cross-disease attention module further pushes the result, which obtains 1.5\% Ac and 0.6\% AUC gain over Zoom-in-net.
For DME grading, our model excels the other reported results~\cite{al2016diabetic} by 2.4\% improvement on Ac metric.

\begin{figure*}[t]
\centering
\includegraphics[width=1.0\textwidth]{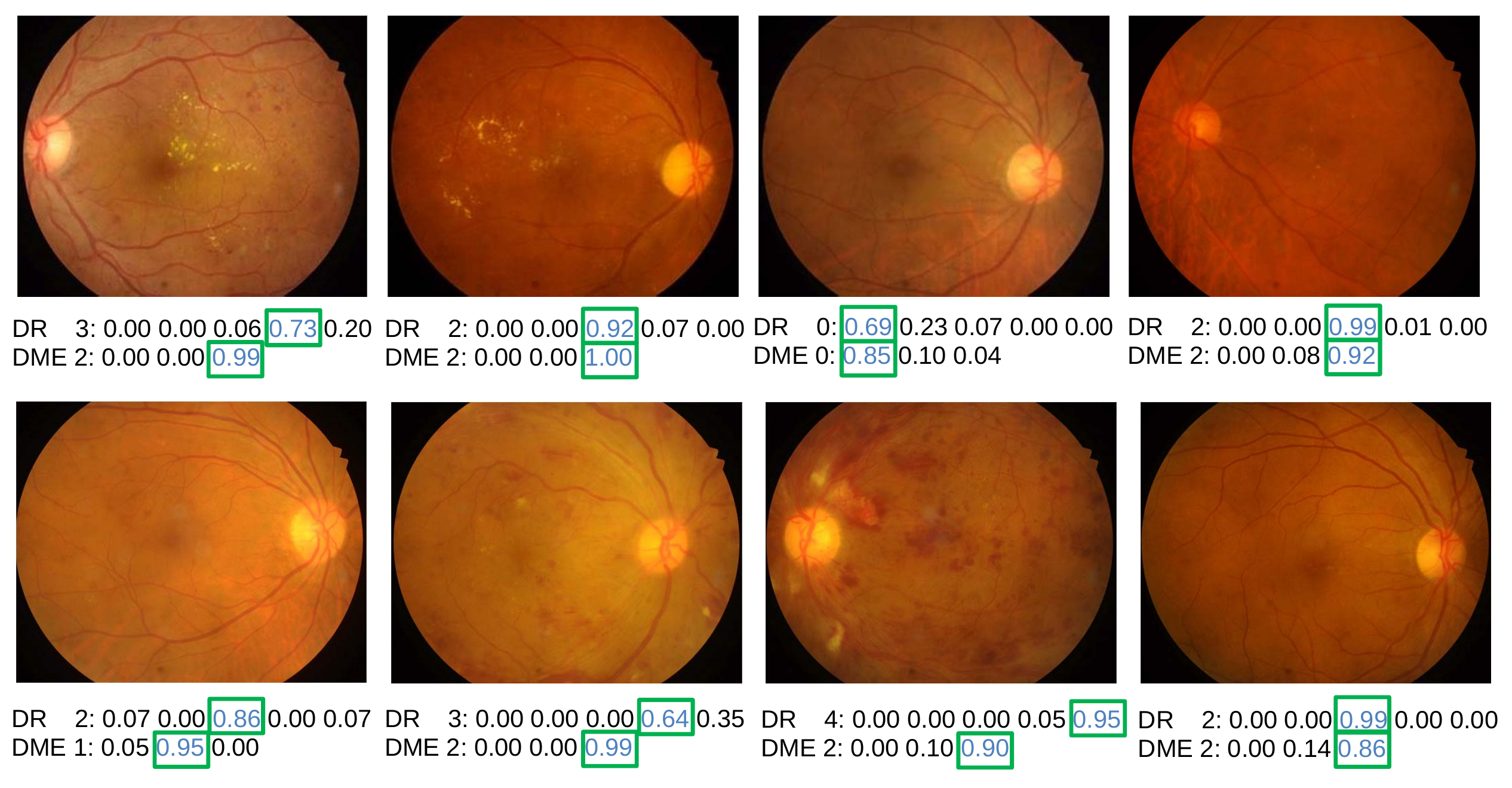}
\caption{Visual results of our method on the test set in the IDRiD dataset. We list the ground-truth, followed by the prediction score for different severity for each individual disease in a sequential order (0-4 for DR and 0-2 for DME). Blue indicates our predicted grade and green box indicates the ground-truth. }
\label{fig:result}
\end{figure*}


\subsection{Results on the IDRiD Challenge Leaderboard}
\begin{table}[h]
\centering
\caption{Comparison with the reported results on the IDRiD leaderboard. (unit: \%)}
{\resizebox{0.55\linewidth}{!}	{\begin{tabular}{c|c|c}
			\toprule[1.5pt]
			Methods  &  Joint Ac  &  Rank
			\tabularnewline  \hline
			\textbf{CANet (ours)} &  \textbf{65.1} & \textbf{1} \tabularnewline
			lzyuncc  & 63.1 &  2 \tabularnewline
			VRT  &  55.3 &  3  \tabularnewline
			
			Mammoth  &  51.5 &  4  \tabularnewline
			
			HarangiM1  & 47.6  &  5  \tabularnewline
			
			AVSASVA & 47.6  & 5  \tabularnewline
			
			HarangiM2  & 40.8 &  6  \tabularnewline
			
			\bottomrule[1.5pt]	
		\end{tabular}}
		\label{tab:resultIDRiD}}
\end{table}
Table~\ref{tab:resultIDRiD} shows the results of our method and other challenge participation methods  on the IDRiD challenge dataset.\footnote{Challenge results are in https://idrid.grand-challenge.org/Leaderboard/}
Our model is trained with only the data in the Sub-challenge 2 (image-level supervision).
It is observed that our model achieves a joint accuracy of 65.1\%, which is higher than the top-ranked result by LzyUNCC (an unpublished work) on the leaderboard, with a relative 2.0\% improvement on the joint accuracy.
Lastly, it is worth noting that we trained our model using only the data in Sub-challenge 2 in the IDRiD dataset, while others (unpublished works) may use model ensembles or other supervision provided in other Sub-challenges.

\begin{figure}
	\centering
	\subfloat[]{{\includegraphics[width=4cm]{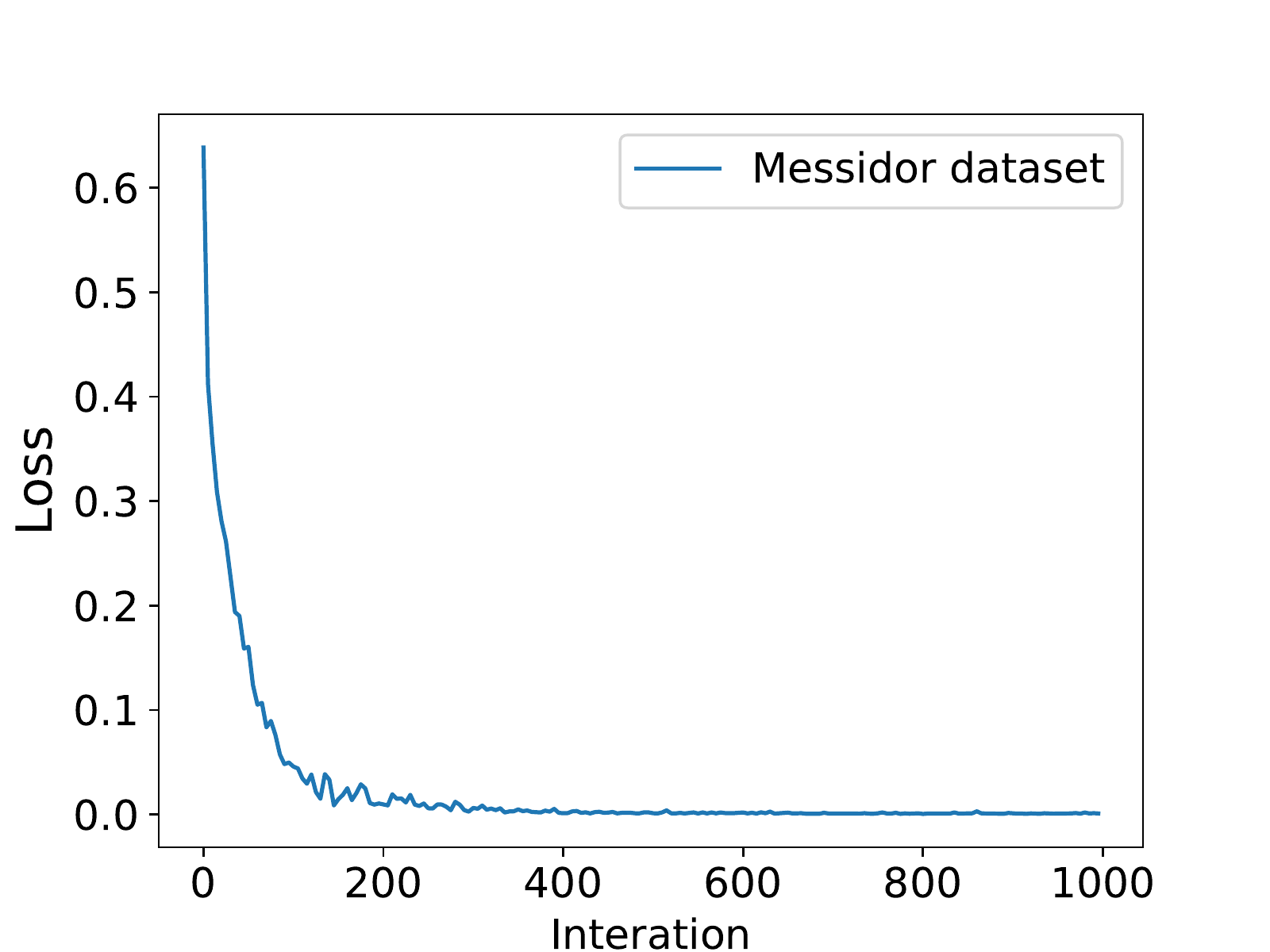}}}
	\qquad
	\subfloat[ ]{{\includegraphics[width=4cm]{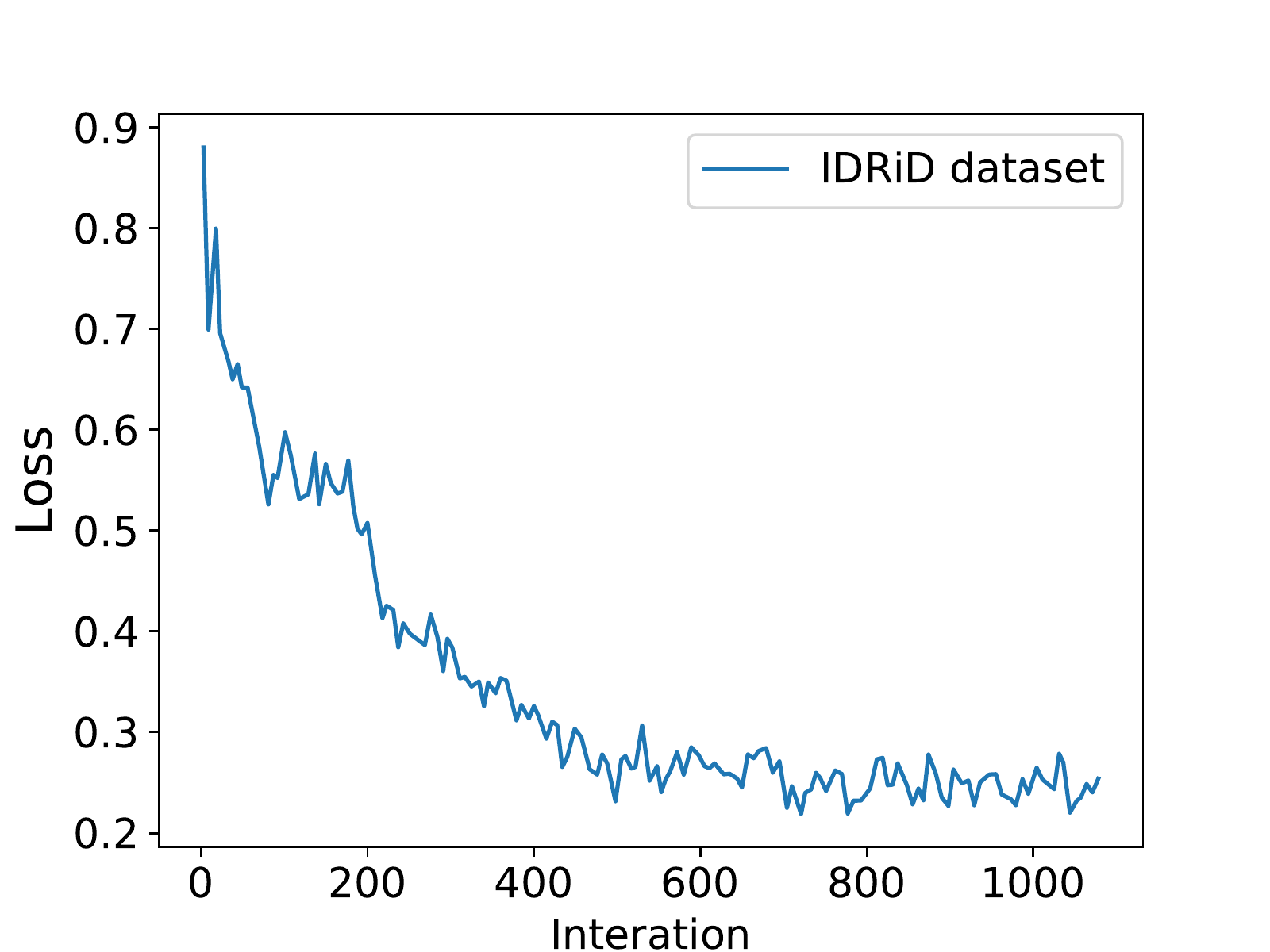} }}
	\caption{\revisesecond{The learning curves of our method on the Messidor dataset (a) and IDRiD dataset (b).}}
	\label{fig:fig_loss}
\end{figure}

We also analyze the effect of each attention module on the IDRiD dataset, and the results are shown in Table~\ref{tab:ablationIDRiD}.
With only disease-specific attention modules (CANet (d-S only)), our method excels the joint training baseline by 1\%.
Two disease-dependent modules ``CANet (d-S, d-D {\tiny{DR$\Rightarrow$DME}})'' and ``CANet (d-S, d-D {\tiny{DR$\Leftarrow$DME}})'' both further improve the joint grading performance by exploring the dependence between these two diseases.
We can also observe that DME has much influences for the grading of DR, and this observation is consistent with that in the Messidor dataset in Table~\ref{tab:ablationstudy}.
With both direction dependent attention modules, our method achieves the best performance with Joint Ac 65.1\%.
Finally, we visualize some examples of the disease prediction score of our method on the IDRiD dataset in Figure~\ref{fig:result}.
We can see that our method clearly differentiates the severity for DR and DME, respectively.
\revisesecond{As shown in Figure~\ref{fig:fig_loss}, we visualize the learning curves of our method on the Messidor dataset and IDRiD dataset, respectively. }
\begin{table}[tp]
	\centering
	\caption{Results on the IDRiD dataset with different attention modules setting. (unit: \%)}
	{\resizebox{0.65\linewidth}{!}{\begin{tabular}{c|c}
	\toprule[1.5pt]
	Methods &    Joint Ac
	\tabularnewline  \hline
	Joint training  & 59.2 \tabularnewline
	CANet (d-S only)  & 60.2 \tabularnewline
	CANet (d-S, d-D {\tiny{DR$\Rightarrow$DME}} ) & 62.1 \tabularnewline
	CANet (d-S, d-D {\tiny{DR$\Leftarrow$DME}} ) & 63.1 \tabularnewline
	\textbf{CANet (final model)}  & \textbf{65.1} \tabularnewline
	\bottomrule[1.5pt]	
\end{tabular}}}
\label{tab:ablationIDRiD}
\end{table}

%% file: discussion.tex
\section{Discussion}

Recently, with the advances of deep learning techniques, automatic grading of DR and DME has been widely studied in the research  community~\cite{ren2018diabetic,syed2018fundus,acharya2017automated,sahlsten2019deep,ramachandran2018diabetic,zhou2019collaborative,lin2018framework,chen2018diabetic,wang2017zoom}. 
Although the large improvements have been achieved on these tasks, to the best of our knowledge, there is no previous works that jointly grade these two diseases and model the relationship between them.
In this work, we investigate the importance of the relationship between DR and DME for the joint grading task, and propose a cross-disease attention network (CANet) to capture the relationship between these two diseases.  
One method consists of two kinds of attention modules: one to learn disease-specific features and another to learn disease-dependent features.
Results shown on two public benchmark datasets,~\ie, the Messidor dataset and 2018 ISBI IDRiD dataset, demonstrated the effectiveness of our method.

Although the good performance achieves, the limitation of our method still exists. 
The whole network is trained with only image-level supervision, making it very challenging to find the accurate abnormal signs, such as soft exudates, hard exudates, microaneurysms, and hemorrhage. 
The lesion masks or bounding boxes would provide the location information of these abnormal signs, which would be largely beneficial to the grading tasks~\cite{lin2018framework,zhou2019collaborative,orlando2018ensemble}, since the severity is usually based on the lesions. 
However, we are not aware of any public datasets containing both DR, DME grading labels, as well as the lesion or abnormal region segmentation masks. 
One solution is to collect the datasets with massive annotations,~\ie, lesion masks and the grading labels of multi-diseases.
Another feasible solution is to explore how to utilize the lesion segmentation information from additional datasets to help the joint DR and DME grading.
The dataset with lesion masks and dataset with DR \& DME grading labels may have domain shifts, and generative adversarial networks~\cite{zhao2016energy,ledig2017photo,arjovsky2017wasserstein,shen2019patient} will be beneficial for this task.

\revise{Our method is feasible to extend to more correlated diseases. 
The attention mechanism aims to learn the attentional weights among multiple diseases. 
If we have multiple correlated diseases, the architecture will have multiple outputs, and each of them is optimized by an individual loss function to obtain the disease-specific features.
Moreover, the disease-dependent attention module can be added to these diseases. 
For example, if there are five correlated diseases, 20 disease-dependent attention modules should be designed, and each module learns the correlation between every two diseases. Due to the high computational cost, the limitations would be the effective design of such attention blocks.}

The future direction we would like to work on is to better model the relationship between DR and its complication DME, and also explore the relationship of multi-diseases occurred in one image. One potential research direction is to use the graph convolutional neural network~\cite{zhou2018graph} to model the relationship among different diseases. Through this, we hope to leverage the correlative information to improve joint grading performance. Also, it might bring some new insights to help doctors in understanding the diseases and their complications, even in a board area of AI medicine~\cite{zhao2019incorporating,cao2019fully}.

%% file: conclusion.tex
\section{Conclusion}
In this work, we present a cross-disease attention network (CANet) to jointly grade DR and DME, 
and explore the individual diseases and also the internal relationship between two diseases by formulating two attention modules: one to learn disease-specific features and another to learn disease-dependent features. 
After that, the network leverages these two features simultaneously for DR and DME grading to maximize the overall grading performance. 
Experimental results on the public Messidor dataset demonstrate the superiority of our network over other related methods on both the DR and DME grading tasks.
Moreover, our method also achieves the best results on the IDRiD challenge dataset.
In the future, we plan to train our network jointly with the lesion annotations to further improve the DR and DME grading performance.

%% file: AAAAmain.bbl
\begin{thebibliography}{10}
\providecommand{\url}[1]{#1}
\csname url@samestyle\endcsname
\providecommand{\newblock}{\relax}
\providecommand{\bibinfo}[2]{#2}
\providecommand{\BIBentrySTDinterwordspacing}{\spaceskip=0pt\relax}
\providecommand{\BIBentryALTinterwordstretchfactor}{4}
\providecommand{\BIBentryALTinterwordspacing}{\spaceskip=\fontdimen2\font plus
\BIBentryALTinterwordstretchfactor\fontdimen3\font minus
  \fontdimen4\font\relax}
\providecommand{\BIBforeignlanguage}[2]{{%
\expandafter\ifx\csname l@#1\endcsname\relax
\typeout{** WARNING: IEEEtran.bst: No hyphenation pattern has been}%
\typeout{** loaded for the language `#1'. Using the pattern for}%
\typeout{** the default language instead.}%
\else
\language=\csname l@#1\endcsname
\fi
#2}}
\providecommand{\BIBdecl}{\relax}
\BIBdecl

\bibitem{cho2018idf}
N.~Cho, J.~Shaw, S.~Karuranga, Y.~Huang, J.~da~Rocha~Fernandes, A.~Ohlrogge
  \emph{et~al.}, ``Idf diabetes atlas: Global estimates of diabetes prevalence
  for 2017 and projections for 2045,'' \emph{Diabetes research and clinical
  practice}, vol. 138, pp. 271--281, 2018.

\bibitem{das2015diabetic}
A.~Das, P.~G. McGuire, and S.~Rangasamy, ``Diabetic macular edema:
  pathophysiology and novel therapeutic targets,'' \emph{Ophthalmology}, vol.
  122, no.~7, pp. 1375--1394, 2015.

\bibitem{islam2018deep}
S.~M.~S. Islam, M.~M. Hasan, and S.~Abdullah, ``Deep learning based early
  detection and grading of diabetic retinopathy using retinal fundus images,''
  \emph{International Conference on Machine Learning, Image Processing, Network
  Security and Data Sciences}, 2018.

\bibitem{kang2018multi}
K.~Zhou, Z.~Gu, W.~Liu, W.~Luo, J.~Cheng, S.~Gao \emph{et~al.}, ``Multi-cell
  multi-task convolutional neural networks for diabetic retinopathy grading,''
  in \emph{International Conference of the IEEE Engineering in Medicine and
  Biology Society}.\hskip 1em plus 0.5em minus 0.4em\relax IEEE, 2018, pp.
  2724--2727.

\bibitem{krause2018grader}
J.~Krause, V.~Gulshan, E.~Rahimy, P.~Karth, K.~Widner, G.~S. Corrado
  \emph{et~al.}, ``Grader variability and the importance of reference standards
  for evaluating machine learning models for diabetic retinopathy,''
  \emph{Ophthalmology}, vol. 125, no.~8, pp. 1264--1272, 2018.

\bibitem{ren2018diabetic}
F.~Ren, P.~Cao, D.~Zhao, and C.~Wan, ``Diabetic macular edema grading in
  retinal images using vector quantization and semi-supervised learning,''
  \emph{Technology and Health Care}, no. Preprint, pp. 1--9, 2018.

\bibitem{syed2018fundus}
A.~M. Syed, M.~U. Akram, T.~Akram, M.~Muzammal, S.~Khalid, and M.~A. Khan,
  ``Fundus images-based detection and grading of macular edema using robust
  macula localization,'' \emph{IEEE Access}, vol.~6, pp. 58\,784--58\,793,
  2018.

\bibitem{gulshan2016development}
V.~Gulshan, L.~Peng, M.~Coram, M.~C. Stumpe, D.~Wu, A.~Narayanaswamy
  \emph{et~al.}, ``Development and validation of a deep learning algorithm for
  detection of diabetic retinopathy in retinal fundus photographs,''
  \emph{Jama}, vol. 316, no.~22, pp. 2402--2410, 2016.

\bibitem{chen2018multi}
Q.~Chen, Y.~Peng, T.~Keenan, S.~Dharssi, E.~Agro, W.~T. Wong \emph{et~al.}, ``A
  multi-task deep learning model for the classification of age-related macular
  degeneration,'' \emph{AMIA Summits on Translational Science Proceedings},
  vol. 2019, p. 505, 2019.

\bibitem{tan2018deep}
C.~Tan, L.~Zhao, Z.~Yan, K.~Li, D.~Metaxas, and Y.~Zhan, ``Deep multi-task and
  task-specific feature learning network for robust shape preserved organ
  segmentation,'' in \emph{2018 IEEE 15th International Symposium on Biomedical
  Imaging (ISBI 2018)}.\hskip 1em plus 0.5em minus 0.4em\relax IEEE, 2018, pp.
  1221--1224.

\bibitem{liu2018mtmr}
L.~Liu, Q.~Dou, H.~Chen, I.~E. Olatunji, J.~Qin, and P.-A. Heng, ``Mtmr-net:
  Multi-task deep learning with margin ranking loss for lung nodule analysis,''
  in \emph{Deep Learning in Medical Image Analysis and Multimodal Learning for
  Clinical Decision Support}.\hskip 1em plus 0.5em minus 0.4em\relax Springer,
  2018, pp. 74--82.

\bibitem{porwal2018indian}
P.~Porwal, S.~Pachade, R.~Kamble, M.~Kokare, G.~Deshmukh, V.~Sahasrabuddhe
  \emph{et~al.}, ``Indian diabetic retinopathy image dataset (idrid): A
  database for diabetic retinopathy screening research,'' \emph{Data}, vol.~3,
  no.~3, p.~25, 2018.

\bibitem{decenciere2014feedback}
E.~Decenci{\`e}re, X.~Zhang, G.~Cazuguel, B.~Lay, B.~Cochener, C.~Trone
  \emph{et~al.}, ``Feedback on a publicly distributed image database: the
  messidor database,'' \emph{Image Analysis \& Stereology}, vol.~33, no.~3, pp.
  231--234, 2014.

\bibitem{silberman2010case}
N.~Silberman, K.~Ahrlich, R.~Fergus, and L.~Subramanian, ``Case for automated
  detection of diabetic retinopathy,'' in \emph{AAAI Spring Symposium Series},
  2010.

\bibitem{sopharak2009automatic}
A.~Sopharak, B.~Uyyanonvara, and S.~Barman, ``Automatic exudate detection from
  non-dilated diabetic retinopathy retinal images using fuzzy c-means
  clustering,'' \emph{sensors}, vol.~9, no.~3, pp. 2148--2161, 2009.

\bibitem{roychowdhury2013dream}
S.~Roychowdhury, D.~D. Koozekanani, and K.~K. Parhi, ``Dream: diabetic
  retinopathy analysis using machine learning,'' \emph{IEEE journal of
  biomedical and health informatics}, vol.~18, no.~5, pp. 1717--1728, 2013.

\bibitem{acharya2009computer}
U.~R. Acharya, C.~M. Lim, E.~Y.~K. Ng, C.~Chee, and T.~Tamura, ``Computer-based
  detection of diabetes retinopathy stages using digital fundus images,''
  \emph{Proceedings of the institution of mechanical engineers, part H: journal
  of engineering in medicine}, vol. 223, no.~5, pp. 545--553, 2009.

\bibitem{akram2014detection}
M.~U. Akram, S.~Khalid, A.~Tariq, S.~A. Khan, and F.~Azam, ``Detection and
  classification of retinal lesions for grading of diabetic retinopathy,''
  \emph{Computers in biology and medicine}, vol.~45, pp. 161--171, 2014.

\bibitem{antal2014ensemble}
B.~Antal and A.~Hajdu, ``An ensemble-based system for automatic screening of
  diabetic retinopathy,'' \emph{Knowledge-based systems}, vol.~60, pp. 20--27,
  2014.

\bibitem{seoud2015red}
L.~Seoud, T.~Hurtut, J.~Chelbi, F.~Cheriet, and J.~P. Langlois, ``Red lesion
  detection using dynamic shape features for diabetic retinopathy screening,''
  \emph{IEEE transactions on medical imaging}, vol.~35, no.~4, pp. 1116--1126,
  2015.

\bibitem{kumar2017kernel}
N.~Kumar, A.~V. Rajwade, S.~Chandran, and S.~P. Awate, ``Kernel
  generalized-gaussian mixture model for robust abnormality detection,'' in
  \emph{MICCAI}.\hskip 1em plus 0.5em minus 0.4em\relax Springer, 2017, pp.
  21--29.

\bibitem{pratt2016convolutional}
H.~Pratt, F.~Coenen, D.~M. Broadbent, S.~P. Harding, and Y.~Zheng,
  ``Convolutional neural networks for diabetic retinopathy,'' \emph{Procedia
  Computer Science}, vol.~90, pp. 200--205, 2016.

\bibitem{li2017convolutional}
X.~Li, T.~Pang, B.~Xiong, W.~Liu, P.~Liang, and T.~Wang, ``Convolutional neural
  networks based transfer learning for diabetic retinopathy fundus image
  classification,'' in \emph{2017 10th International Congress on Image and
  Signal Processing, BioMedical Engineering and Informatics (CISP-BMEI)}.\hskip
  1em plus 0.5em minus 0.4em\relax IEEE, 2017, pp. 1--11.

\bibitem{kori2018ensemble}
A.~Kori, S.~S. Chennamsetty, M.~S. K.P., and V.~Alex, ``Ensemble of
  convolutional neural networks for automatic grading of diabetic retinopathy
  and macular edema,'' \emph{arXiv preprint arXiv:1809.04228}, 2018.

\bibitem{xiao2019major}
D.~Xiao, A.~Bhuiyan, S.~Frost, J.~Vignarajan, M.-L. Tay-Kearney, and
  Y.~Kanagasingam, ``Major automatic diabetic retinopathy screening systems and
  related core algorithms: a review,'' \emph{Machine Vision and Applications},
  vol.~30, no.~3, pp. 423--446, 2019.

\bibitem{cao2018efficient}
P.~Cao, F.~Ren, C.~Wan, J.~Yang, and O.~Zaiane, ``Efficient multi-kernel
  multi-instance learning using weakly supervised and imbalanced data for
  diabetic retinopathy diagnosis,'' \emph{Computerized Medical Imaging and
  Graphics}, vol.~69, pp. 112--124, 2018.

\bibitem{hagos2019transfer}
M.~T. Hagos and S.~Kant, ``Transfer learning based detection of diabetic
  retinopathy from small dataset,'' \emph{arXiv preprint arXiv:1905.07203},
  2019.

\bibitem{jama2019}
D.~S.~W. Ting, L.~Carin, and M.~D. Abramoff, ``{Observations and Lessons
  Learned From the Artificial Intelligence Studies for Diabetic Retinopathy
  Screening},'' \emph{JAMA Ophthalmology}, 06 2019.

\bibitem{van2016fast}
M.~J. Van~Grinsven, B.~van Ginneken, C.~B. Hoyng, T.~Theelen, and C.~I.
  S{\'a}nchez, ``Fast convolutional neural network training using selective
  data sampling: Application to hemorrhage detection in color fundus images,''
  \emph{IEEE transactions on medical imaging}, vol.~35, no.~5, pp. 1273--1284,
  2016.

\bibitem{dai2017retinal}
L.~Dai, B.~Sheng, Q.~Wu, H.~Li, X.~Hou, W.~Jia \emph{et~al.}, ``Retinal
  microaneurysm detection using clinical report guided multi-sieving cnn,'' in
  \emph{MICCAI}.\hskip 1em plus 0.5em minus 0.4em\relax Springer, 2017, pp.
  525--532.

\bibitem{yang2017lesion}
Y.~Yang, T.~Li, W.~Li, H.~Wu, W.~Fan, and W.~Zhang, ``Lesion detection and
  grading of diabetic retinopathy via two-stages deep convolutional neural
  networks,'' in \emph{MICCAI}.\hskip 1em plus 0.5em minus 0.4em\relax
  Springer, 2017, pp. 533--540.

\bibitem{lin2018framework}
Z.~Lin, R.~Guo, Y.~Wang, B.~Wu, T.~Chen, W.~Wang \emph{et~al.}, ``A framework
  for identifying diabetic retinopathy based on anti-noise detection and
  attention-based fusion,'' in \emph{MICCAI}.\hskip 1em plus 0.5em minus
  0.4em\relax Springer, 2018, pp. 74--82.

\bibitem{zhou2019collaborative}
Y.~Zhou, X.~He, L.~Huang, L.~Liu, F.~Zhu, S.~Cui \emph{et~al.}, ``Collaborative
  learning of semi-supervised segmentation and classification for medical
  images,'' in \emph{CVPR}, 2019, pp. 2079--2088.

\bibitem{gargeya2017automated}
R.~Gargeya and T.~Leng, ``Automated identification of diabetic retinopathy
  using deep learning,'' \emph{Ophthalmology}, vol. 124, no.~7, pp. 962--969,
  2017.

\bibitem{wang2017zoom}
Z.~Wang, Y.~Yin, J.~Shi, W.~Fang, H.~Li, and X.~Wang, ``Zoom-in-net: Deep
  mining lesions for diabetic retinopathy detection,'' in \emph{MICCAI}.\hskip
  1em plus 0.5em minus 0.4em\relax Springer, 2017, pp. 267--275.

\bibitem{fraz2018computational}
M.~Fraz, M.~Badar, A.~Malik, and S.~Barman, ``Computational methods for
  exudates detection and macular edema estimation in retinal images: a
  survey,'' \emph{Archives of Computational Methods in Engineering}, pp. 1--28,
  2018.

\bibitem{akram2014automated}
M.~U. Akram, A.~Tariq, S.~A. Khan, and M.~Y. Javed, ``Automated detection of
  exudates and macula for grading of diabetic macular edema,'' \emph{Computer
  methods and programs in biomedicine}, vol. 114, no.~2, pp. 141--152, 2014.

\bibitem{acharya2017automated}
U.~R. Acharya, M.~R.~K. Mookiah, J.~E. Koh, J.~H. Tan, S.~V. Bhandary, A.~K.
  Rao \emph{et~al.}, ``Automated diabetic macular edema (dme) grading system
  using dwt, dct features and maculopathy index,'' \emph{Computers in biology
  and medicine}, vol.~84, pp. 59--68, 2017.

\bibitem{al2016diabetic}
B.~Al-Bander, W.~Al-Nuaimy, M.~A. Al-Taee, B.~M. Williams, and Y.~Zheng,
  ``Diabetic macular edema grading based on deep neural networks.''\hskip 1em
  plus 0.5em minus 0.4em\relax University of Iowa, 2016.

\bibitem{moeskops2016deep}
P.~Moeskops, J.~M. Wolterink, B.~H. van~der Velden, K.~G. Gilhuijs, T.~Leiner,
  M.~A. Viergever \emph{et~al.}, ``Deep learning for multi-task medical image
  segmentation in multiple modalities,'' in \emph{MICCAI}, 2016, pp. 478--486.

\bibitem{xue2018full}
W.~Xue, G.~Brahm, S.~Pandey, S.~Leung, and S.~Li, ``Full left ventricle
  quantification via deep multitask relationships learning,'' \emph{Medical
  Image Analysis}, vol.~43, pp. 54--65, 2018.

\bibitem{he2016deep}
K.~He, X.~Zhang, S.~Ren, and J.~Sun, ``Deep residual learning for image
  recognition,'' in \emph{CVPR}, 2016.

\bibitem{woo2018cbam}
S.~Woo, J.~Park, J.-Y. Lee, and I.~So~Kweon, ``Cbam: Convolutional block
  attention module,'' in \emph{ECCV}, 2018, pp. 3--19.

\bibitem{hu2018squeeze}
J.~Hu, L.~Shen, and G.~Sun, ``Squeeze-and-excitation networks,'' in
  \emph{CVPR}, 2018, pp. 7132--7141.

\bibitem{kingma2014adam}
D.~P. Kingma and J.~Ba, ``Adam: A method for stochastic optimization,'' in
  \emph{ICLR}, 2015.

\bibitem{loshchilov2016sgdr}
I.~Loshchilov and F.~Hutter, ``Sgdr: Stochastic gradient descent with warm
  restarts,'' in \emph{ICLR}, 2017.

\bibitem{paszke2017automatic}
A.~Paszke, S.~Gross, S.~Chintala, G.~Chanan, E.~Yang, Z.~DeVito \emph{et~al.},
  ``Automatic differentiation in pytorch,'' 2017.

\bibitem{sanchez2011evaluation}
C.~I. S{\'a}nchez, M.~Niemeijer, A.~V. Dumitrescu, M.~S. Suttorp-Schulten,
  M.~D. Abramoff, and B.~van Ginneken, ``Evaluation of a computer-aided
  diagnosis system for diabetic retinopathy screening on public data,''
  \emph{Investigative ophthalmology \& visual science}, vol.~52, no.~7, pp.
  4866--4871, 2011.

\bibitem{vo2016new}
H.~H. Vo and A.~Verma, ``New deep neural nets for fine-grained diabetic
  retinopathy recognition on hybrid color space,'' in \emph{IEEE International
  Symposium on Multimedia}.\hskip 1em plus 0.5em minus 0.4em\relax IEEE, 2016,
  pp. 209--215.

\bibitem{pires2015beyond}
R.~Pires, S.~Avila, H.~F. Jelinek, J.~Wainer, E.~Valle, and A.~Rocha, ``Beyond
  lesion-based diabetic retinopathy: a direct approach for referral,''
  \emph{IEEE journal of biomedical and health informatics}, vol.~21, no.~1, pp.
  193--200, 2015.

\bibitem{huang2017densely}
G.~Huang, Z.~Liu, L.~Van Der~Maaten, and K.~Q. Weinberger, ``Densely connected
  convolutional networks,'' in \emph{CVPR}, 2017, pp. 4700--4708.

\bibitem{cuadros2009eyepacs}
``Kaggle diabetic retinopathy detection competition,''
  \url{https://www.kaggle.com/c/diabetic-retinopathy-detection}.

\bibitem{ramachandran2018diabetic}
N.~Ramachandran, S.~C. Hong, M.~J. Sime, and G.~A. Wilson, ``Diabetic
  retinopathy screening using deep neural network,'' \emph{Clinical \&
  experimental ophthalmology}, vol.~46, no.~4, pp. 412--416, 2018.

\bibitem{sahlsten2019deep}
J.~Sahlsten, J.~Jaskari, J.~Kivinen, L.~Turunen, E.~Jaanio, K.~Hietala
  \emph{et~al.}, ``Deep learning fundus image analysis for diabetic retinopathy
  and macular edema grading,'' \emph{arXiv preprint arXiv:1904.08764}, 2019.

\bibitem{chen2018diabetic}
Y.-W. Chen, T.-Y. Wu, W.-H. Wong, and C.-Y. Lee, ``Diabetic retinopathy
  detection based on deep convolutional neural networks,'' in \emph{2018 IEEE
  International Conference on Acoustics, Speech and Signal Processing
  (ICASSP)}.\hskip 1em plus 0.5em minus 0.4em\relax IEEE, 2018, pp. 1030--1034.

\bibitem{orlando2018ensemble}
J.~I. Orlando, E.~Prokofyeva, M.~del Fresno, and M.~B. Blaschko, ``An ensemble
  deep learning based approach for red lesion detection in fundus images,''
  \emph{Computer methods and programs in biomedicine}, vol. 153, pp. 115--127,
  2018.

\bibitem{zhao2016energy}
J.~Zhao, M.~Mathieu, and Y.~LeCun, ``Energy-based generative adversarial
  network,'' \emph{arXiv preprint arXiv:1609.03126}, 2016.

\bibitem{ledig2017photo}
C.~Ledig, L.~Theis, F.~Husz{\'a}r, J.~Caballero, A.~Cunningham, A.~Acosta
  \emph{et~al.}, ``Photo-realistic single image super-resolution using a
  generative adversarial network,'' in \emph{CVPR}, 2017, pp. 4681--4690.

\bibitem{arjovsky2017wasserstein}
M.~Arjovsky, S.~Chintala, and L.~Bottou, ``Wasserstein generative adversarial
  networks,'' in \emph{ICML}, 2017, pp. 214--223.

\bibitem{shen2019patient}
L.~Shen, W.~Zhao, and L.~Xing, ``Patient-specific reconstruction of volumetric
  computed tomography images from a single projection view via deep learning,''
  \emph{Nature biomedical engineering}, pp. 1--9, 2019.

\bibitem{zhou2018graph}
J.~Zhou, G.~Cui, Z.~Zhang, C.~Yang, Z.~Liu, and M.~Sun, ``Graph neural
  networks: A review of methods and applications,'' \emph{arXiv preprint
  arXiv:1812.08434}, 2018.

\bibitem{zhao2019incorporating}
W.~Zhao, B.~Han, Y.~Yang, M.~Buyyounouski, S.~L. Hancock, H.~Bagshaw, and
  L.~Xing, ``Incorporating imaging information from deep neural network layers
  into image guided radiation therapy (igrt),'' \emph{Radiotherapy and
  Oncology}, vol. 140, pp. 167--174, 2019.

\bibitem{cao2019fully}
L.~Cao, R.~Shi, Y.~Ge, L.~Xing, P.~Zuo, Y.~Jia, J.~Liu, Y.~He, X.~Wang, S.~Luan
  \emph{et~al.}, ``Fully automatic segmentation of type b aortic dissection
  from cta images enabled by deep learning,'' \emph{European Journal of
  Radiology}, 2019.

\end{thebibliography}
